\documentclass[%
 reprint,
 amsmath,amssymb,
 aps,prapplied,
]{revtex4-2}

\usepackage{graphicx}% Include figure files
\usepackage{dcolumn}% Align table columns on decimal point
\usepackage{bm}% bold math
\usepackage{xcolor}
\begin{document}

%\preprint{}

\title{High performance cryogen-free microkelvin platform}


\author{J.~Ny\'eki}
\author{M.~Lucas}
\author{P.~Knappov\'a}
\author{L.V.~Levitin}
\author{A.~Casey}
\email{a.casey@rhul.ac.uk, author to whom correspondence should be addressed}
\author{J.~Saunders}
%\thanks{}
\affiliation{Department of Physics, Royal Holloway, University of London, Egham,
Surrey, TW20 0EX, UK}
\author{H.~van~der~Vliet}
\author{A.J.~Matthews}
\affiliation{Oxford Instruments NanoScience, Tubney Woods, Abingdon,
Oxfordshire, OX13 5QX, UK}


\date{\today}

\begin{abstract}
Improved accessibility to the microkelvin temperature regime is important for future research in quantum materials; for quantum information science; and for applications of quantum sensors. Here we report the design and performance of a microkelvin platform based on a nuclear demagnetization stage, engineered and well optimized for operation on a standard cryogen-free dilution refrigerator. PrNi$_5$ is used as the dominant refrigerant. The platform provides a large area for mounting experiments in an ultralow temperature, low electromagnetic noise environment. The performance is characterized using current sensing noise thermometry. Temperatures as low as 395~$\mu$K have been reached, and a protocol has been established in which it is possible to operate experiments below 1~mK for $95\%$ of the time, providing an efficient cryogen-free microkelvin environment for a wide range of science applications.
% We demonstrate cooling to 395~$\mu$K. We have established a mode of operation where it is possible to be below 1~mK for $95\%$ of the time, providing an efficient cryogen-free microkelvin environment for a wide range of science applications.

\end{abstract}

%\keywords{}

\maketitle

\section{Background}
Advances in cryogen-free dilution refrigeration \cite{Uhlig2002,BATEY2009} have played a crucial role in the development of superconducting quantum technology for computing and sensing. The microkelvin temperature regime represents a strategically important frontier for quantum information science. Cryogen-free technology is essential to promote the accessibility of this regime and to ensure sustainability, beyond specialist infrastructures such as the European Microkelvin Platform \footnote{{h}ttps://emplatform.eu/}. Cooling into this regime is achieved by adiabatic nuclear demagnetization \cite{Pobell2007}. The nuclear spins of a suitable material are polarized in an external magnetic field and cooled to dilution refrigerator temperatures. After thermally isolating the material, the external field is then reduced. Under approaching isentropic conditions, in the absence of interactions, $T_f=T_{i} B_{f}/B_{i}$, where $T_{f/i}, B_{f/i}$ refer to the final and initial temperature and magnetic field.

The range of quantum materials and quantum sensors of interest for quantum information science is extremely diverse. Access to lower temperatures will yield fundamental insights into two-level fluctuators in superconducting qubits \cite{deGraaf2017,deGraaf2020}, which are a barrier to fault-tolerant quantum computing \cite{Shor1996,Muller2019}. Further cooling of two dimensional electron gases \cite{Samkharadze2011}, will promote understanding of aspects of the fractional quantum Hall effect \cite{Pan1999} and drive the development of semiconductor devices, including spintronic devices \cite{Ho2018,Hays2021}. There is an ongoing search for a crystalline topological superconductor \cite{Sato_2017,Shaffer2020,Sharma2020, Nayak2021, Nogaki2021, Khim2021} and YbRh$_2$Si$_2$ \cite{Schuberth2016} is a candidate material that requires temperatures below a few mK. When cooling below 2.5~mK $^3$He provides a benchmark for topological superconductivity, displaying a wealth of phenomena in the surface and edge states that arise from bulk-surface correspondence \cite{Mizushima2016, Makinen2019,Volovik2021, Heikkinen2021}. Strongly correlated electron systems exhibit quantum criticality \cite{Nakatsuji2008,Fuhrman2021, Paschen2021}, which is essentially a $T=0$ phenomenon. An exciting direction is the extension of studies of quantum criticality to lower temperature, for example to understand the role of nuclear magnetism, particularly in systems with strong hyperfine interactions \cite{Chekhovich2013, Loss2018, Kikkawa2021,Eisenlohr2021}. Cooling to below 1~mK allows access to magnetic field-to-temperature ratios, $B/T$, unachievable elsewhere. High $B/T$ ratios are of particular interest in heavy fermion systems and also to investigate hyperfine effects. Finally microkelvin platforms will be of increasing importance for hosting quantum sensors for fundamental physics, such as dark matter searches and quantum simulators. These may require locating in underground facilities, where cryogen-free operation is optimal.

Significant effort has been devoted to on-chip cooling technologies \cite{Muhonen_2012, Pekola2021}. Two successful approaches to reaching microkelvin temperatures have been either by parallel nuclear demagnetization of each lead connected to a sample \cite{Clark2010, Palma2017} or by the incorporation of nuclear refrigeration elements directly on-chip, where the device is itself a thermometer (a Coulomb blockade thermometer), which directly measured the electron temperature \cite{Jones2020-ie,Sarsby2020}.

Our approach is to treat the cooling of the microkelvin platform and the thermalization of the device/sample to this platform separately. The merit of this approach is its flexibility and applicability to a wide range of materials and devices. On traditional ``wet-cryostats'' we have demonstrated: cooling of a two dimensional electron gas below 1~mK in a $^3$He immersion cell \cite{Levitin2022}; thermodynamic and transport measurements on a heavy fermion superconductor YbRh$_2$Si$_2$ down to 200~$\mu$K \cite{Knapp2022}; and an ac susceptibility study of superconductivity and antiferroquadrupole order in PrOs$_4$Sb$_{12}$ \cite{Bangma2022} down to 1 mK, reaching a $B/T$ in excess of 6000~T/K. In combination with cryogen-free operation this opens a wide landscape for research on quantum materials into the microkelvin regime.
\section{Introduction}
A \emph{proof-of-principle} cryogen-free nuclear demagnetization platform \cite{Batey_2013} was based upon a PrNi$_5$ nuclear stage, that had been originally constructed for a traditional ``wet-system'' \cite{Parpia1985}. This demonstrated that the vibrations inherent in a pulse-tube cooled cryostat \cite{SCHMORANZER2019102} are not incompatible with nuclear demagnetization. Cryogen-free systems with copper nuclear demagnetization stages have been developed reaching temperatures below 100~$\mu$K \cite{Todoshchenko2014,Palma2017, Yan2021}. Designs of continuous cryogen-free nuclear demagnetization systems have been proposed \cite{Toda_2018, SCHMORANZER2020} but have yet to have demonstrated their performance.

The focus of this paper is a cryogen-free microkelvin platform capable of achieving temperatures below 500~$\mu$K with a hold time below 1~mK suitable for the applications discussed previously. A rapid thermal turnaround time dramatically enhances the efficiency of an experiment, with a greater fraction of time being spent in the temperature range of interest. A key figure of merit is the thermal ``duty cycle'', which we define as the percentage of the total time of the single-shot cooling cycle spent below 1~mK. For our design the best value achieved so far is $95\%$, a dramatic improvement over the previous published results for cryogen-free systems, ranging from $50\%$ down to $< 10\%$ \cite{Batey_2013,Todoshchenko2014,Yan2021}.

To achieve this result we combined the favorable properties of the Van Vleck paramagnetic material PrNi$_5$, which has previously been used in traditional ``wet-cryostats'' \cite{Andres1977,MUELLER1980,Folle1981,Greywall1985,WIEGERS1990, Pobell2007}, with a well designed thermal link, heat switch and system to minimize the relative motion between the demagnetization magnet and nuclear stage.

Essential to the characterization of the performance of the system is reliable thermometry. We adopted current sensing noise thermometry (CSNT) \cite{Lusher2000, Shibahara2016}, which can be used over the full temperature range.
\section{Design Details}
The design principle was for a system that would be compatible with a minimally modified cryogen-free dilution refrigerator, boosting the lowest temperatures that could be reached to below 500~$\mu$K (in this case an Oxford Instruments Triton 200 \footnote{Oxford Instruments NanoScience, UK. https://www.oxinst.com/}). The system was equipped with a 6~T superconducting magnet. Field compensation results in less than 10~mT in the ultralow temperature (ULT) experimental region, when the magnet is at the full field. We adopted a modular design in which the thermal links interconnecting the main components of the platform used demountable cone joints \cite{Rybalko1996}.

\begin{figure}
	\includegraphics[width=\columnwidth]{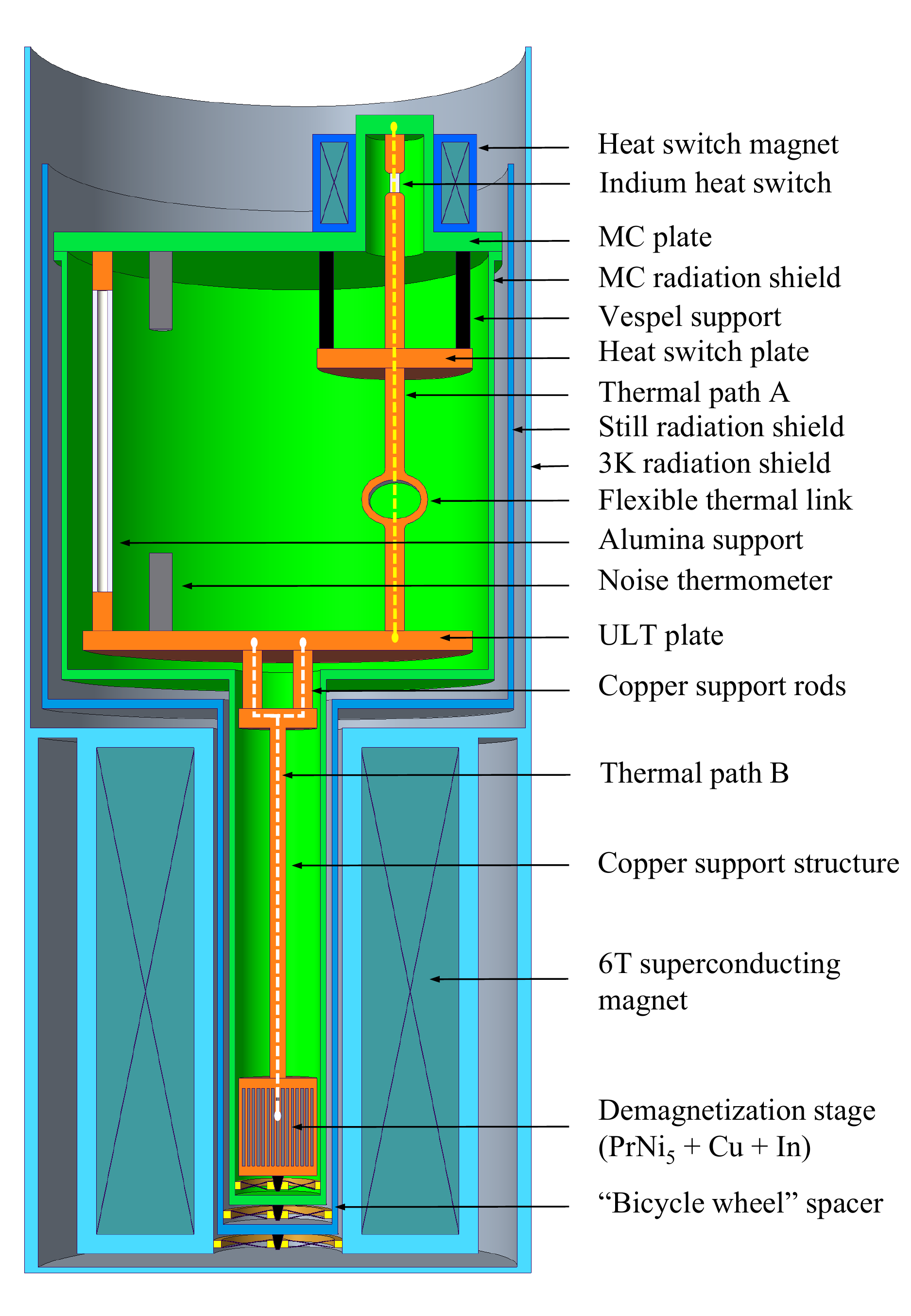}
	\caption{\label{fig1} Schematic diagram of the microkelvin platform. Enclosed in green is the ultralow temperature, low electrical noise region. The current sensing noise thermometers on the ULT and MC plates are shown in grey. The thermal pathways for the link from the mixing chamber to the ULT plate and for the ULT plate to the PrNi$_5$ are marked by dashed lines.}
\end{figure}

A schematic of the system is shown in Fig.\ref{fig1}. All of the copper used in the construction was commercial OFE-OK copper \footnote{Luvata Wolverhampton Ltd., UK.\newline https://www.luvata.com/}. Through heat treatment the residual resistivity ratio (RRR) of the copper could be increased up to 3000. The mixing chamber (MC) plate, the heat switch (HS) plate, and the ULT plate were gold-plated to prevent oxidization of the surfaces \footnote{Twickenham Plating Group, UK.\newline http://twickenham.co.uk/}. Each plate was equipped with a 4-wire resistive heater for characterization. Simplicity and reproducibility of construction, and the avoidance of hazardous materials and toxic chemicals in the construction process were also design criteria. 

The original cryostat MC plate was replaced by a design to meet the requirement for an RF-tight space for the microkelvin platform. The MC shield and plate enclose a large volume, creating an electrically quiet environment for measurements while reducing the heat leak to the ULT region. The HS magnet is mounted on the MC plate, outside of the RF shield. A CSNT (sensor resistance 300~m$\Omega$) was mounted on the MC plate. 

The experimental volume contains a relatively large area ULT plate, of diameter 240~mm, with a RRR of 265. The ULT plate is equipped with multiple mounting points for experiments, consisting of 14 cone joints, 3 of which have line-of-sight access to the room temperature ports on the top of the cryostat. The ULT plate is suspended from the MC plate via alumina tubes. The thermal pathway to the MC plate is via a flexible copper link (annealed to achieve a RRR of 3000), the HS plate and the robust indium heat switch. The HS plate is suspended from the MC plate using Vespel rods. A CSNT (sensor resistance 2~m$\Omega$) was mounted on the ULT plate and was used to characterize the performance of the demagnetization process.

The nuclear demagnetization stage was constructed out of polycrystalline PrNi$_5$ \footnote{AMES Laboratory, U.S. Department of Energy, Iowa State University} of total mass 94.5~gm (0.22~mole), supplied in the form of irregular rods, soldered using indium to a rigid copper support structure. The copper support structure (RRR of 1660) is attached to the ULT plate by copper rods with cone joints at each end. While all of the construction materials contribute to nuclear cooling, the dominant contribution arises from the PrNi$_5$. Hyperfine enhancement boosts the field seen by the $^{141}$Pr nuclei by a factor 11.2 above the externally applied field \cite{Kubota1980}. In 6~T a significant entropy reduction ($>80\%$) of the PrNi$_5$ can be achieved in the modest temperature range, 10~-~20~mK. 

Praseodymium's small Korringa constant \cite{KORRINGA1950}, $\kappa$, results in a negligible thermal resistance between the electrons and the nuclear spin system ($\kappa_{\mathrm{Pr}} \sim$~3~$\mu$Ks \cite{Herrmannsdorfer1994} compared to $\kappa_{\mathrm{Cu}}=$~1.2~Ks for copper \cite{Pobell2007}). The thermal conductivity of polycrystalline PrNi$_5$ is poor, comparable to brass. The PrNi$_5$ rods had a RRR that varied from 7 to 30. It is therefore critical to optimize the heat transfer between the copper support structure and the PrNi$_5$. This was achieved using indium as a solder \cite{WIEGERS1990}. Cadmium has traditionally been used for this purpose \cite{MUELLER1980} but has higher toxicity, zinc has recently been proposed \cite{Takimoto2022} as an alternative material. Although indium limits the final field to be in excess of 28~mT, the superconducting critical field of indium, it did not compromise the achievement of the performance objectives.

To achieve a short precooling time, the thermal link between the MC plate and the nuclear stage was designed to be dominant relative to the cooling power of the MC. For comparison at the end of the precool shown in figure \ref{fig2}, the cooling power of the dilution refrigerator was $\dot{Q}=5~\mu\rm{W}$, resulting in $\dot{Q}/T^{2}\sim 12.2 ~\rm{mW/K^{2}}$, less than half of the value $L/2R=27~\rm{mW/K^{2}}$ due to the thermal link, where $L$ is the Lorenz number \cite{lorenz1872} and $R\sim 450 ~\rm{n\Omega}$ is the total resistance between the PrNi$_5$ and the MC. Figure \ref{fig1} shows this thermal pathway in two sections, \emph{Path A}, between the MC and ULT plate, and \emph{Path B}, between the ULT plate and the Pr nuclear spin bath. The measured thermal resistance of \emph{Path A}, with the heat switch in its normal state, corresponds, via the Weidemann-Franz law \cite{Franz1853}, to an electrical resistance of 300~n$\Omega$. The superconducting heat switch assembly including cone joints contributes 170~n$\Omega$ to this value. The thermal conductivity of the open switch, measured at 100~mK is 3$\times10^{-5}$~Wcm$^{-1}$K$^{-1}$, equivalent to that achieved in aluminum heat switches \cite{Pobell2007, Butterworth2022}. The thermal resistance of \emph{Path B} was measured directly, by applying heat to the ULT plate, and corresponded to an electrical resistance of 150~n$\Omega$.

Key to the cooling performance achieved is the immunity to mechanical vibrations. It is most important to limit relative motion of the nuclear stage and the 6~T magnet. This is achieved by three similar “bicycle-wheel” spacers: between stage and MC shield; between MC shield and still shield; and between the still shield and main magnet (see Fig.1 and Fig.\ref{fig5} inset). 

\section{Performance}
\begin{figure}
	\includegraphics[width=\columnwidth]{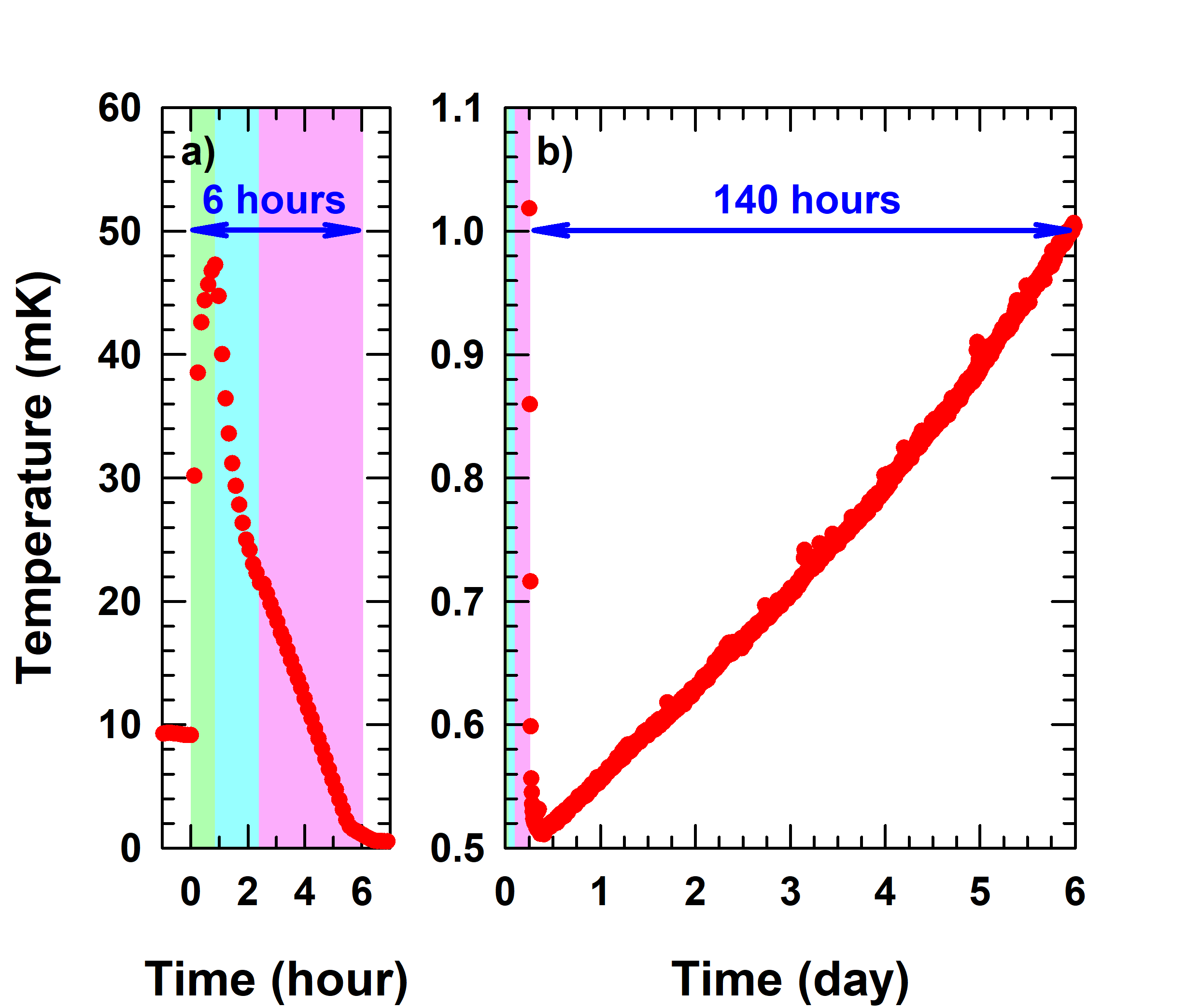}
	\caption{\label{fig2} a) Preparation time for a ``lunchtime'' precool; the external magnetic field is ramped up to 6~T (green); the magnet is persisted and the nuclear stage is allowed to pre-cool (blue); the heat switch is opened and the nuclear stage is demagnetized to 30~mT (pink). b) Temperature of the ULT plate CSNT as a function of time following the demagnetization, showing a hold-time of 140~hours below 1~mK.}
\end{figure}

To illustrate the duty cycle we show in Fig.\ref{fig2} a process in which the system starts in a low field at 9~mK, it is magnetized to 6~T, followed by a rapid pre-cool period of about 2~hours. We refer to this as a “lunchtime” precool. After opening the superconducting heat switch, a demagnetization to a final field of 30~mT at a constant rate of 1.75~T/hour gives a final temperature around 500~$\mu$K. The heat leak due to eddy currents was measured to be $30~\rm{nW}$ in the field interval 5.6~T to 4.5~T. The platform remains below 1~mK for almost 6~days. With longer pre-cool time (15~hours, referred to as an ``overnight'' precool), the starting temperature of the demagnetization is typically 10~mK and with a staggered demagnetization rate over a period of 7.5~hours a temperature of 395~$\mu$K was achieved. Eddy current heating in the copper structure, arising from vibrations while pre-cooling in the high magnetic field and sweeping of the field during demagnetization, is tolerable despite its geometry and high RRR. Here we exploit the high magneto-resistance of copper \cite{Arenz1982, Fickett1988}; with a RRR~=~1660 the electrical resistance increases by a factor of over 20 in 6~T.

To quantify the heat leak and its field dependence the total heat capacity of the nuclear stage was measured using the CSNT and a heater on the ULT plate. We exploit the achievable precision in short measurement times, $t_m$, using two-stage SQUID detection \cite{Drung2007}. The noise floor of these devices allow the sensor resistor to be from a few m$\Omega$ to a few hundred m$\Omega$ and still have a reasonable noise temperature (in the range 5~$\mu$K to 100~$\mu$K). The 2 m$\Omega$ sensor of the CSNT delivers 1\% precision in $t_m$=10~s \cite{Lusher2000}.

\begin{figure}
	\includegraphics[width=\columnwidth]{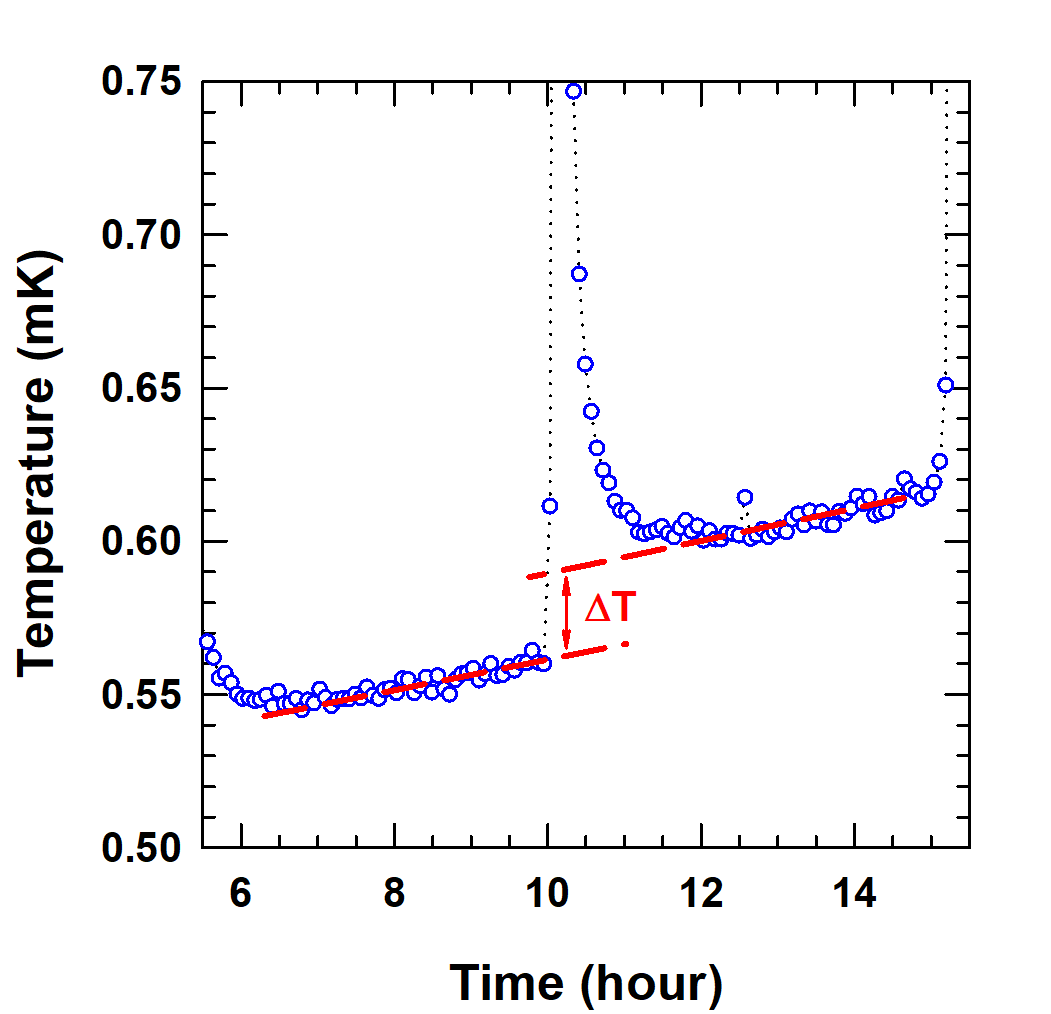}
	\caption{\label{fig3} A typical heat pulse measurement to evaluate the heat capacity ($\sim$~70 nW applied for 10 minutes), demonstrating that a $\Delta T$ corresponding to 5$\%$ temperature step can easily be resolved at temperatures as low as 0.5~mK.}
\end{figure}

\begin{figure}
	\includegraphics[width=\columnwidth]{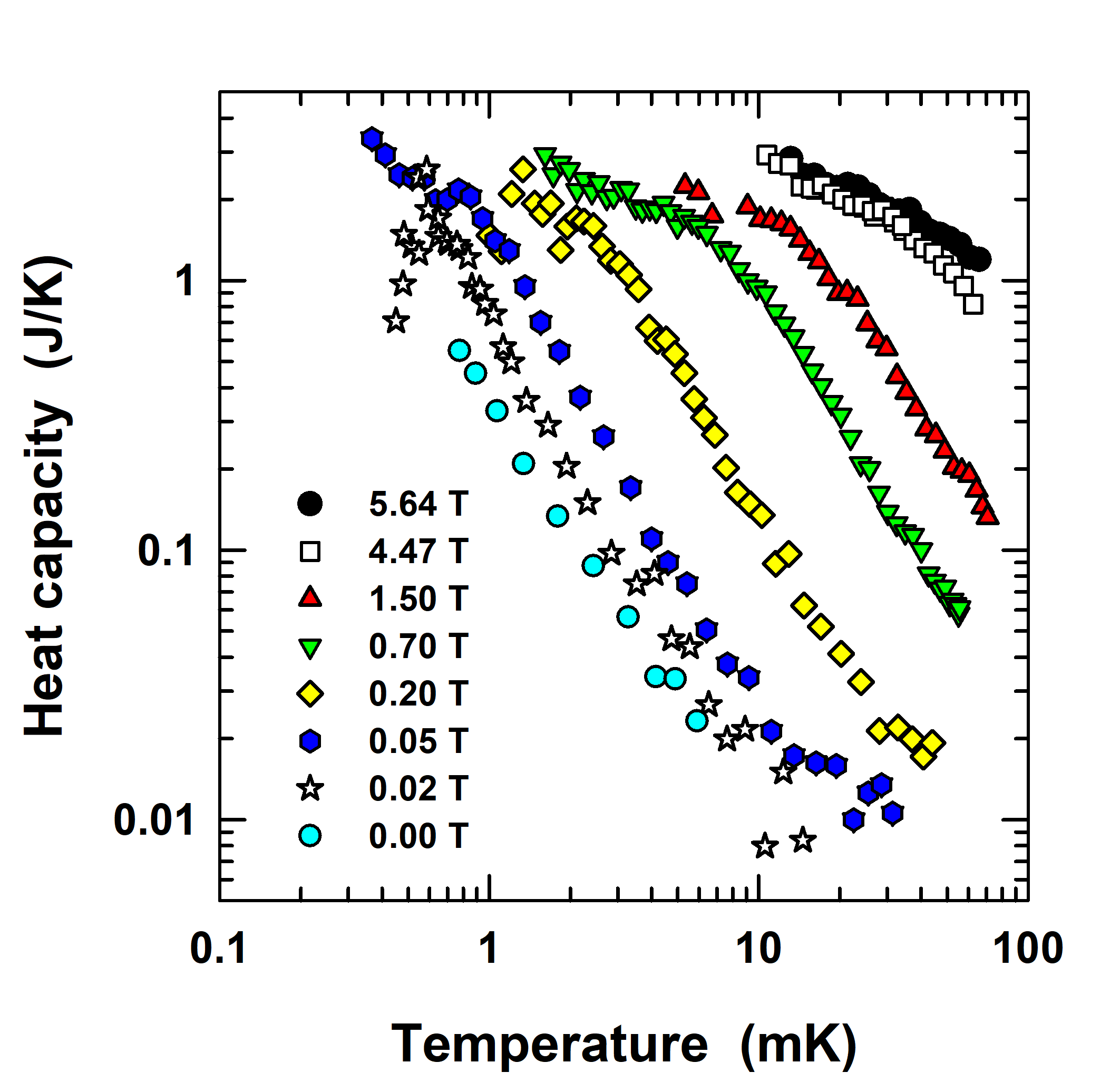}
	\caption{\label{fig4} Total heat capacity of the nuclear stage as a function of temperature and applied magnetic field, with PrNi$_5$ providing the dominant contribution.}
\end{figure}

\begin{figure}
	\includegraphics[width=\columnwidth]{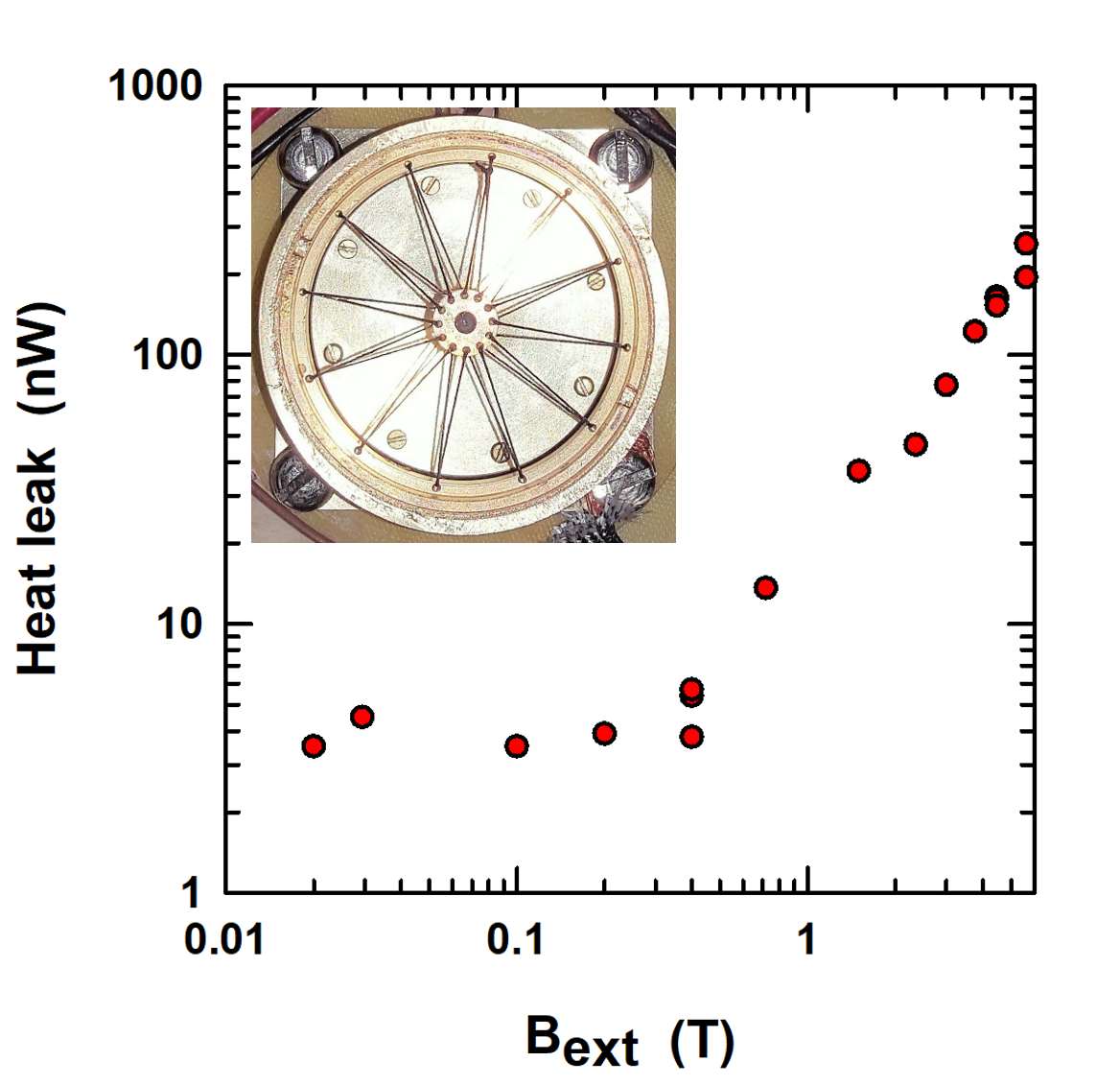}
	\caption{\label{fig5} The heat leak to the nuclear demagnetization stage as a function of external magnetic field, extracted from warming data. Subsequent improvements resulted in a 30~mT heat leak of 2~nW, which is the heat leak under which the warming data shown in Fig.\ref{fig2} was obtained. Inset: ``Bicycle-wheel'' components connecting the still shield to the magnet.}
\end{figure}

A typical heat pulse is shown in Fig.\ref{fig3}. The total stage heat capacity measured in this way for various magnetic fields is shown in Fig.\ref{fig4}. The residual heat leak in different magnetic fields is determined by simply observing the warm up rate, Fig.\ref{fig5}. We note that with this level of heat leak in low fields, and the experimentally determined thermal resistance between the ULT plate and the PrNi$_5$, we can be confident that the CSNT is in good equilibrium with the stage. Thermalization of the CSNT to the ULT plate relies on filtering the leads between the SQUID and the sensor resistor \cite{CSNT2021}.

The 4~nW field independent heat leak at low fields Fig.\ref{fig5} was subsequently reduced to 2~nW, by improvements in electrical and mechanical grounding of room temperature services. The warm-up data in Fig.\ref{fig2} was determined under these conditions, and we estimate that the ``overnight'' pre-cool will result in two weeks below 1~mK with such a low heat leak.
\section{Conclusion}
We have demonstrated the performance of a microkelvin platform, that is modular, robust, and tolerant to mechanical vibrations. The platform can rapidly reach temperatures as low as 395~$\mu$K. The holdtime below 1~mK represents an excellent duty cycle of $95\%$.

The platform provides a large electrically quiet experimental microkelvin volume, capable of hosting multiple experiments, including those requiring line-of-sight access to room temperature.

The modularity of the design ensures compatibility for upgrading existing cryogen-free dilution refrigerators, offering the opportunity for an order of magnitude decrease in their base temperature. To promote future sustainability and reduce the environmental impact the manufacturing processes were designed to minimize the use of toxic and hazardous materials. The technological importance of PrNi$_5$ as a working material motivates the growth of samples of improved quality.

This work moves us from a \emph{proof-of-principle} to a prototype system, with very limited demands on the surrounding infrastructure. The internal levels of vibration isolation achieved were sufficient to produce a low heat leak to the demagnetization stage, without massive support structures and optical-table quality air-mounts. As such this is a significant step towards the accessibility of microkelvin temperatures, opening up a frontier in quantum information science and quantum material research to the widest possible community.
\begin{acknowledgments}
We would like to thank Mark Meisel for useful discussions. The work in the London Low Temperature Laboratory was supported by the technical staff, in particular Richard Elsom, Ian Higgs, Paul Bamford, and Harpal Sandhu. The research leading to these results has received funding from the European Union’s Horizon 2020 Research and Innovation Programme, under Grant Agreement no. 824109.

\end{acknowledgments}

\bibliography{DryDemagbib.bib}

\begin{thebibliography}{69}%
\makeatletter
\providecommand \@ifxundefined [1]{%
 \@ifx{#1\undefined}
}%
\providecommand \@ifnum [1]{%
 \ifnum #1\expandafter \@firstoftwo
 \else \expandafter \@secondoftwo
 \fi
}%
\providecommand \@ifx [1]{%
 \ifx #1\expandafter \@firstoftwo
 \else \expandafter \@secondoftwo
 \fi
}%
\providecommand \natexlab [1]{#1}%
\providecommand \enquote  [1]{``#1''}%
\providecommand \bibnamefont  [1]{#1}%
\providecommand \bibfnamefont [1]{#1}%
\providecommand \citenamefont [1]{#1}%
\providecommand \href@noop [0]{\@secondoftwo}%
\providecommand \href [0]{\begingroup \@sanitize@url \@href}%
\providecommand \@href[1]{\@@startlink{#1}\@@href}%
\providecommand \@@href[1]{\endgroup#1\@@endlink}%
\providecommand \@sanitize@url [0]{\catcode `\\12\catcode `\$12\catcode
  `\&12\catcode `\#12\catcode `\^12\catcode `\_12\catcode `\%12\relax}%
\providecommand \@@startlink[1]{}%
\providecommand \@@endlink[0]{}%
\providecommand \url  [0]{\begingroup\@sanitize@url \@url }%
\providecommand \@url [1]{\endgroup\@href {#1}{\urlprefix }}%
\providecommand \urlprefix  [0]{URL }%
\providecommand \Eprint [0]{\href }%
\providecommand \doibase [0]{https://doi.org/}%
\providecommand \selectlanguage [0]{\@gobble}%
\providecommand \bibinfo  [0]{\@secondoftwo}%
\providecommand \bibfield  [0]{\@secondoftwo}%
\providecommand \translation [1]{[#1]}%
\providecommand \BibitemOpen [0]{}%
\providecommand \bibitemStop [0]{}%
\providecommand \bibitemNoStop [0]{.\EOS\space}%
\providecommand \EOS [0]{\spacefactor3000\relax}%
\providecommand \BibitemShut  [1]{\csname bibitem#1\endcsname}%
\let\auto@bib@innerbib\@empty
%</preamble>
\bibitem [{\citenamefont {Uhlig}(2002)}]{Uhlig2002}%
  \BibitemOpen
  \bibfield  {author} {\bibinfo {author} {\bibfnamefont {K.}~\bibnamefont
  {Uhlig}},\ }\bibfield  {title} {\bibinfo {title}
  {$^{3}\mathrm{He}/^{4}\mathrm{He}$ dilution refrigerator with pulse-tube
  refrigerator precooling},\ }\href@noop {} {\bibfield  {journal} {\bibinfo
  {journal} {Cryogenics}\ }\textbf {\bibinfo {volume} {42}},\ \bibinfo {pages}
  {73} (\bibinfo {year} {2002})}\BibitemShut {NoStop}%
\bibitem [{\citenamefont {Batey}\ \emph {et~al.}(2009)\citenamefont {Batey},
  \citenamefont {Buehler}, \citenamefont {Cuthbert}, \citenamefont {Foster},
  \citenamefont {Matthews}, \citenamefont {Teleberg},\ and\ \citenamefont
  {Twin}}]{BATEY2009}%
  \BibitemOpen
  \bibfield  {author} {\bibinfo {author} {\bibfnamefont {G.}~\bibnamefont
  {Batey}}, \bibinfo {author} {\bibfnamefont {M.}~\bibnamefont {Buehler}},
  \bibinfo {author} {\bibfnamefont {M.}~\bibnamefont {Cuthbert}}, \bibinfo
  {author} {\bibfnamefont {T.}~\bibnamefont {Foster}}, \bibinfo {author}
  {\bibfnamefont {A.}~\bibnamefont {Matthews}}, \bibinfo {author}
  {\bibfnamefont {G.}~\bibnamefont {Teleberg}},\ and\ \bibinfo {author}
  {\bibfnamefont {A.}~\bibnamefont {Twin}},\ }\bibfield  {title} {\bibinfo
  {title} {Integration of superconducting magnets with cryogen-free dilution
  refrigerator systems},\ }\href@noop {} {\bibfield  {journal} {\bibinfo
  {journal} {Cryogenics}\ }\textbf {\bibinfo {volume} {49}},\ \bibinfo {pages}
  {727} (\bibinfo {year} {2009})}\BibitemShut {NoStop}%
\bibitem [{Note1()}]{Note1}%
  \BibitemOpen
  \bibinfo {note} {{h}ttps://emplatform.eu/}\BibitemShut {NoStop}%
\bibitem [{\citenamefont {Pobell}(2007)}]{Pobell2007}%
  \BibitemOpen
  \bibfield  {author} {\bibinfo {author} {\bibfnamefont {F.}~\bibnamefont
  {Pobell}},\ }\href@noop {} {\emph {\bibinfo {title} {Matter and methods at
  low temperatures}}}\ (\bibinfo  {publisher} {Springer Berlin},\ \bibinfo
  {year} {2007})\ pp.\ \bibinfo {pages} {1--461}\BibitemShut {NoStop}%
\bibitem [{\citenamefont {de~Graaf}\ \emph {et~al.}(2017)\citenamefont
  {de~Graaf}, \citenamefont {Adamyan}, \citenamefont {Lindstr\"om},
  \citenamefont {Erts}, \citenamefont {Kubatkin}, \citenamefont {Tzalenchuk},\
  and\ \citenamefont {Danilov}}]{deGraaf2017}%
  \BibitemOpen
  \bibfield  {author} {\bibinfo {author} {\bibfnamefont {S.~E.}\ \bibnamefont
  {de~Graaf}}, \bibinfo {author} {\bibfnamefont {A.~A.}\ \bibnamefont
  {Adamyan}}, \bibinfo {author} {\bibfnamefont {T.}~\bibnamefont
  {Lindstr\"om}}, \bibinfo {author} {\bibfnamefont {D.}~\bibnamefont {Erts}},
  \bibinfo {author} {\bibfnamefont {S.~E.}\ \bibnamefont {Kubatkin}}, \bibinfo
  {author} {\bibfnamefont {A.~Y.}\ \bibnamefont {Tzalenchuk}},\ and\ \bibinfo
  {author} {\bibfnamefont {A.~V.}\ \bibnamefont {Danilov}},\ }\bibfield
  {title} {\bibinfo {title} {Direct identification of dilute surface spins on
  ${\mathrm{al}}_{2}{\mathrm{o}}_{3}$: {O}rigin of flux noise in quantum
  circuits},\ }\href {https://doi.org/10.1103/PhysRevLett.118.057703}
  {\bibfield  {journal} {\bibinfo  {journal} {Phys. Rev. Lett.}\ }\textbf
  {\bibinfo {volume} {118}},\ \bibinfo {pages} {057703} (\bibinfo {year}
  {2017})}\BibitemShut {NoStop}%
\bibitem [{\citenamefont {de~Graaf}\ \emph {et~al.}(2020)\citenamefont
  {de~Graaf}, \citenamefont {Faoro}, \citenamefont {Ioffe}, \citenamefont
  {Mahashabde}, \citenamefont {Burnett}, \citenamefont {Lindström},
  \citenamefont {Kubatkin}, \citenamefont {Danilov},\ and\ \citenamefont
  {Tzalenchuk}}]{deGraaf2020}%
  \BibitemOpen
  \bibfield  {author} {\bibinfo {author} {\bibfnamefont {S.~E.}\ \bibnamefont
  {de~Graaf}}, \bibinfo {author} {\bibfnamefont {L.}~\bibnamefont {Faoro}},
  \bibinfo {author} {\bibfnamefont {L.~B.}\ \bibnamefont {Ioffe}}, \bibinfo
  {author} {\bibfnamefont {S.}~\bibnamefont {Mahashabde}}, \bibinfo {author}
  {\bibfnamefont {J.~J.}\ \bibnamefont {Burnett}}, \bibinfo {author}
  {\bibfnamefont {T.}~\bibnamefont {Lindström}}, \bibinfo {author}
  {\bibfnamefont {S.~E.}\ \bibnamefont {Kubatkin}}, \bibinfo {author}
  {\bibfnamefont {A.~V.}\ \bibnamefont {Danilov}},\ and\ \bibinfo {author}
  {\bibfnamefont {A.~Y.}\ \bibnamefont {Tzalenchuk}},\ }\bibfield  {title}
  {\bibinfo {title} {Two-level systems in superconducting quantum devices due
  to trapped quasiparticles},\ }\href@noop {} {\bibfield  {journal} {\bibinfo
  {journal} {Science Advances}\ }\textbf {\bibinfo {volume} {6}},\ \bibinfo
  {pages} {eabc5055} (\bibinfo {year} {2020})}\BibitemShut {NoStop}%
\bibitem [{\citenamefont {Shor}(1996)}]{Shor1996}%
  \BibitemOpen
  \bibfield  {author} {\bibinfo {author} {\bibfnamefont {P.}~\bibnamefont
  {Shor}},\ }\bibfield  {title} {\bibinfo {title} {Fault-tolerant quantum
  computation},\ }in\ \href@noop {} {\emph {\bibinfo {booktitle} {Proceedings
  of {37}th Conference on Foundations of Computer Science}}}\ (\bibinfo
  {publisher} {IEEE Computer Society},\ \bibinfo {address} {Los Alamitos, CA,
  USA},\ \bibinfo {year} {1996})\ p.~\bibinfo {pages} {56}\BibitemShut
  {NoStop}%
\bibitem [{\citenamefont {Müller}\ \emph {et~al.}(2019)\citenamefont
  {Müller}, \citenamefont {Cole},\ and\ \citenamefont
  {Lisenfeld}}]{Muller2019}%
  \BibitemOpen
  \bibfield  {author} {\bibinfo {author} {\bibfnamefont {C.}~\bibnamefont
  {Müller}}, \bibinfo {author} {\bibfnamefont {J.~H.}\ \bibnamefont {Cole}},\
  and\ \bibinfo {author} {\bibfnamefont {J.}~\bibnamefont {Lisenfeld}},\
  }\bibfield  {title} {\bibinfo {title} {Towards understanding
  two-level-systems in amorphous solids: insights from quantum circuits},\
  }\href {https://doi.org/10.1088/1361-6633/ab3a7e} {\bibfield  {journal}
  {\bibinfo  {journal} {Reports on Progress in Physics}\ }\textbf {\bibinfo
  {volume} {82}},\ \bibinfo {pages} {124501} (\bibinfo {year}
  {2019})}\BibitemShut {NoStop}%
\bibitem [{\citenamefont {Samkharadze}\ \emph {et~al.}(2011)\citenamefont
  {Samkharadze}, \citenamefont {Kumar}, \citenamefont {Manfra}, \citenamefont
  {Pfeiffer}, \citenamefont {West},\ and\ \citenamefont
  {Csáthy}}]{Samkharadze2011}%
  \BibitemOpen
  \bibfield  {author} {\bibinfo {author} {\bibfnamefont {N.}~\bibnamefont
  {Samkharadze}}, \bibinfo {author} {\bibfnamefont {A.}~\bibnamefont {Kumar}},
  \bibinfo {author} {\bibfnamefont {M.~J.}\ \bibnamefont {Manfra}}, \bibinfo
  {author} {\bibfnamefont {L.~N.}\ \bibnamefont {Pfeiffer}}, \bibinfo {author}
  {\bibfnamefont {K.~W.}\ \bibnamefont {West}},\ and\ \bibinfo {author}
  {\bibfnamefont {G.~A.}\ \bibnamefont {Csáthy}},\ }\bibfield  {title}
  {\bibinfo {title} {Integrated electronic transport and thermometry at
  millikelvin temperatures and in strong magnetic fields},\ }\href@noop {}
  {\bibfield  {journal} {\bibinfo  {journal} {Review of Scientific
  Instruments}\ }\textbf {\bibinfo {volume} {82}},\ \bibinfo {pages} {053902}
  (\bibinfo {year} {2011})}\BibitemShut {NoStop}%
\bibitem [{\citenamefont {Pan}\ \emph {et~al.}(1999)\citenamefont {Pan},
  \citenamefont {Xia}, \citenamefont {Shvarts}, \citenamefont {Adams},
  \citenamefont {Stormer}, \citenamefont {Tsui}, \citenamefont {Pfeiffer},
  \citenamefont {Baldwin},\ and\ \citenamefont {West}}]{Pan1999}%
  \BibitemOpen
  \bibfield  {author} {\bibinfo {author} {\bibfnamefont {W.}~\bibnamefont
  {Pan}}, \bibinfo {author} {\bibfnamefont {J.-S.}\ \bibnamefont {Xia}},
  \bibinfo {author} {\bibfnamefont {V.}~\bibnamefont {Shvarts}}, \bibinfo
  {author} {\bibfnamefont {D.~E.}\ \bibnamefont {Adams}}, \bibinfo {author}
  {\bibfnamefont {H.~L.}\ \bibnamefont {Stormer}}, \bibinfo {author}
  {\bibfnamefont {D.~C.}\ \bibnamefont {Tsui}}, \bibinfo {author}
  {\bibfnamefont {L.~N.}\ \bibnamefont {Pfeiffer}}, \bibinfo {author}
  {\bibfnamefont {K.~W.}\ \bibnamefont {Baldwin}},\ and\ \bibinfo {author}
  {\bibfnamefont {K.~W.}\ \bibnamefont {West}},\ }\bibfield  {title} {\bibinfo
  {title} {Exact quantization of the even-denominator fractional quantum {H}all
  state at {$\nu =5/2$} {L}andau level filling factor},\ }\href@noop {}
  {\bibfield  {journal} {\bibinfo  {journal} {Phys. Rev. Lett.}\ }\textbf
  {\bibinfo {volume} {83}},\ \bibinfo {pages} {3530} (\bibinfo {year}
  {1999})}\BibitemShut {NoStop}%
\bibitem [{\citenamefont {Ho}\ \emph {et~al.}(2018)\citenamefont {Ho},
  \citenamefont {Chang}, \citenamefont {Chang}, \citenamefont {Lo},
  \citenamefont {Creeth}, \citenamefont {Kumar}, \citenamefont {Farrer},
  \citenamefont {Ritchie}, \citenamefont {Griffiths}, \citenamefont {Jones},
  \citenamefont {Pepper},\ and\ \citenamefont {Chen}}]{Ho2018}%
  \BibitemOpen
  \bibfield  {author} {\bibinfo {author} {\bibfnamefont {S.-C.}\ \bibnamefont
  {Ho}}, \bibinfo {author} {\bibfnamefont {H.-J.}\ \bibnamefont {Chang}},
  \bibinfo {author} {\bibfnamefont {C.-H.}\ \bibnamefont {Chang}}, \bibinfo
  {author} {\bibfnamefont {S.-T.}\ \bibnamefont {Lo}}, \bibinfo {author}
  {\bibfnamefont {G.}~\bibnamefont {Creeth}}, \bibinfo {author} {\bibfnamefont
  {S.}~\bibnamefont {Kumar}}, \bibinfo {author} {\bibfnamefont
  {I.}~\bibnamefont {Farrer}}, \bibinfo {author} {\bibfnamefont
  {D.}~\bibnamefont {Ritchie}}, \bibinfo {author} {\bibfnamefont
  {J.}~\bibnamefont {Griffiths}}, \bibinfo {author} {\bibfnamefont
  {G.}~\bibnamefont {Jones}}, \bibinfo {author} {\bibfnamefont
  {M.}~\bibnamefont {Pepper}},\ and\ \bibinfo {author} {\bibfnamefont {T.-M.}\
  \bibnamefont {Chen}},\ }\bibfield  {title} {\bibinfo {title} {Imaging the
  zigzag {W}igner crystal in confinement-tunable quantum wires},\ }\href@noop
  {} {\bibfield  {journal} {\bibinfo  {journal} {Phys. Rev. Lett.}\ }\textbf
  {\bibinfo {volume} {121}},\ \bibinfo {pages} {106801} (\bibinfo {year}
  {2018})}\BibitemShut {NoStop}%
\bibitem [{\citenamefont {Hays}\ \emph {et~al.}(2021)\citenamefont {Hays},
  \citenamefont {Fatemi}, \citenamefont {Bouman}, \citenamefont {Cerrillo},
  \citenamefont {Diamond}, \citenamefont {Serniak}, \citenamefont {Connolly},
  \citenamefont {Krogstrup}, \citenamefont {Nygård}, \citenamefont {Yeyati},
  \citenamefont {Geresdi},\ and\ \citenamefont {Devoret}}]{Hays2021}%
  \BibitemOpen
  \bibfield  {author} {\bibinfo {author} {\bibfnamefont {M.}~\bibnamefont
  {Hays}}, \bibinfo {author} {\bibfnamefont {V.}~\bibnamefont {Fatemi}},
  \bibinfo {author} {\bibfnamefont {D.}~\bibnamefont {Bouman}}, \bibinfo
  {author} {\bibfnamefont {J.}~\bibnamefont {Cerrillo}}, \bibinfo {author}
  {\bibfnamefont {S.}~\bibnamefont {Diamond}}, \bibinfo {author} {\bibfnamefont
  {K.}~\bibnamefont {Serniak}}, \bibinfo {author} {\bibfnamefont
  {T.}~\bibnamefont {Connolly}}, \bibinfo {author} {\bibfnamefont
  {P.}~\bibnamefont {Krogstrup}}, \bibinfo {author} {\bibfnamefont
  {J.}~\bibnamefont {Nygård}}, \bibinfo {author} {\bibfnamefont {A.~L.}\
  \bibnamefont {Yeyati}}, \bibinfo {author} {\bibfnamefont {A.}~\bibnamefont
  {Geresdi}},\ and\ \bibinfo {author} {\bibfnamefont {M.~H.}\ \bibnamefont
  {Devoret}},\ }\bibfield  {title} {\bibinfo {title} {Coherent manipulation of
  an {A}ndreev spin qubit},\ }\href {https://doi.org/10.1126/science.abf0345}
  {\bibfield  {journal} {\bibinfo  {journal} {Science}\ }\textbf {\bibinfo
  {volume} {373}},\ \bibinfo {pages} {430} (\bibinfo {year}
  {2021})}\BibitemShut {NoStop}%
\bibitem [{\citenamefont {Sato}\ and\ \citenamefont {Ando}(2017)}]{Sato_2017}%
  \BibitemOpen
  \bibfield  {author} {\bibinfo {author} {\bibfnamefont {M.}~\bibnamefont
  {Sato}}\ and\ \bibinfo {author} {\bibfnamefont {Y.}~\bibnamefont {Ando}},\
  }\bibfield  {title} {\bibinfo {title} {Topological superconductors: a
  review},\ }\href@noop {} {\bibfield  {journal} {\bibinfo  {journal} {Reports
  on Progress in Physics}\ }\textbf {\bibinfo {volume} {80}},\ \bibinfo {pages}
  {076501} (\bibinfo {year} {2017})}\BibitemShut {NoStop}%
\bibitem [{\citenamefont {Shaffer}\ \emph {et~al.}(2020)\citenamefont
  {Shaffer}, \citenamefont {Kang}, \citenamefont {Burnell},\ and\ \citenamefont
  {Fernandes}}]{Shaffer2020}%
  \BibitemOpen
  \bibfield  {author} {\bibinfo {author} {\bibfnamefont {D.}~\bibnamefont
  {Shaffer}}, \bibinfo {author} {\bibfnamefont {J.}~\bibnamefont {Kang}},
  \bibinfo {author} {\bibfnamefont {F.~J.}\ \bibnamefont {Burnell}},\ and\
  \bibinfo {author} {\bibfnamefont {R.~M.}\ \bibnamefont {Fernandes}},\
  }\bibfield  {title} {\bibinfo {title} {Crystalline nodal topological
  superconductivity and {B}ogolyubov {F}ermi surfaces in monolayer
  {N}b{S}e$_{2}$},\ }\href@noop {} {\bibfield  {journal} {\bibinfo  {journal}
  {Phys. Rev. B}\ }\textbf {\bibinfo {volume} {101}},\ \bibinfo {pages}
  {224503} (\bibinfo {year} {2020})}\BibitemShut {NoStop}%
\bibitem [{\citenamefont {Sharma}\ \emph {et~al.}(2020)\citenamefont {Sharma},
  \citenamefont {Edkins}, \citenamefont {Wang}, \citenamefont {Kostin},
  \citenamefont {Sow}, \citenamefont {Maeno}, \citenamefont {Mackenzie},
  \citenamefont {Davis},\ and\ \citenamefont {Madhavan}}]{Sharma2020}%
  \BibitemOpen
  \bibfield  {author} {\bibinfo {author} {\bibfnamefont {R.}~\bibnamefont
  {Sharma}}, \bibinfo {author} {\bibfnamefont {S.~D.}\ \bibnamefont {Edkins}},
  \bibinfo {author} {\bibfnamefont {Z.}~\bibnamefont {Wang}}, \bibinfo {author}
  {\bibfnamefont {A.}~\bibnamefont {Kostin}}, \bibinfo {author} {\bibfnamefont
  {C.}~\bibnamefont {Sow}}, \bibinfo {author} {\bibfnamefont {Y.}~\bibnamefont
  {Maeno}}, \bibinfo {author} {\bibfnamefont {A.~P.}\ \bibnamefont
  {Mackenzie}}, \bibinfo {author} {\bibfnamefont {J.~C.~S.}\ \bibnamefont
  {Davis}},\ and\ \bibinfo {author} {\bibfnamefont {V.}~\bibnamefont
  {Madhavan}},\ }\bibfield  {title} {\bibinfo {title} {Momentum-resolved
  superconducting energy gaps of {S}r$_{2}${R}u{O}$_{4}$ from quasiparticle
  interference imaging},\ }\href@noop {} {\bibfield  {journal} {\bibinfo
  {journal} {Proceedings of the National Academy of Sciences}\ }\textbf
  {\bibinfo {volume} {117}},\ \bibinfo {pages} {5222} (\bibinfo {year}
  {2020})}\BibitemShut {NoStop}%
\bibitem [{\citenamefont {Nayak}\ \emph {et~al.}(2021)\citenamefont {Nayak},
  \citenamefont {Steinbok}, \citenamefont {Roet}, \citenamefont {Koo},
  \citenamefont {Margalit}, \citenamefont {Feldman}, \citenamefont {Almoalem},
  \citenamefont {Kanigel}, \citenamefont {Fiete}, \citenamefont {Yan},
  \citenamefont {Oreg}, \citenamefont {Avraham},\ and\ \citenamefont
  {Beidenkopf}}]{Nayak2021}%
  \BibitemOpen
  \bibfield  {author} {\bibinfo {author} {\bibfnamefont {A.~K.}\ \bibnamefont
  {Nayak}}, \bibinfo {author} {\bibfnamefont {A.}~\bibnamefont {Steinbok}},
  \bibinfo {author} {\bibfnamefont {Y.}~\bibnamefont {Roet}}, \bibinfo {author}
  {\bibfnamefont {J.}~\bibnamefont {Koo}}, \bibinfo {author} {\bibfnamefont
  {G.}~\bibnamefont {Margalit}}, \bibinfo {author} {\bibfnamefont
  {I.}~\bibnamefont {Feldman}}, \bibinfo {author} {\bibfnamefont
  {A.}~\bibnamefont {Almoalem}}, \bibinfo {author} {\bibfnamefont
  {A.}~\bibnamefont {Kanigel}}, \bibinfo {author} {\bibfnamefont {G.~A.}\
  \bibnamefont {Fiete}}, \bibinfo {author} {\bibfnamefont {B.}~\bibnamefont
  {Yan}}, \bibinfo {author} {\bibfnamefont {Y.}~\bibnamefont {Oreg}}, \bibinfo
  {author} {\bibfnamefont {N.}~\bibnamefont {Avraham}},\ and\ \bibinfo {author}
  {\bibfnamefont {H.}~\bibnamefont {Beidenkopf}},\ }\bibfield  {title}
  {\bibinfo {title} {Evidence of topological boundary modes with topological
  nodal-point superconductivity},\ }\href@noop {} {\bibfield  {journal}
  {\bibinfo  {journal} {Nature Physics}\ }\textbf {\bibinfo {volume} {17}},\
  \bibinfo {pages} {1413} (\bibinfo {year} {2021})}\BibitemShut {NoStop}%
\bibitem [{\citenamefont {Nogaki}\ \emph {et~al.}(2021)\citenamefont {Nogaki},
  \citenamefont {Daido}, \citenamefont {Ishizuka},\ and\ \citenamefont
  {Yanase}}]{Nogaki2021}%
  \BibitemOpen
  \bibfield  {author} {\bibinfo {author} {\bibfnamefont {K.}~\bibnamefont
  {Nogaki}}, \bibinfo {author} {\bibfnamefont {A.}~\bibnamefont {Daido}},
  \bibinfo {author} {\bibfnamefont {J.}~\bibnamefont {Ishizuka}},\ and\
  \bibinfo {author} {\bibfnamefont {Y.}~\bibnamefont {Yanase}},\ }\bibfield
  {title} {\bibinfo {title} {Topological crystalline superconductivity in
  locally noncentrosymmetric {C}e{Rh}$_{2}${A}s$_{2}$},\ }\href@noop {}
  {\bibfield  {journal} {\bibinfo  {journal} {Phys. Rev. Research}\ }\textbf
  {\bibinfo {volume} {3}},\ \bibinfo {pages} {L032071} (\bibinfo {year}
  {2021})}\BibitemShut {NoStop}%
\bibitem [{\citenamefont {Khim}\ \emph {et~al.}(2021)\citenamefont {Khim},
  \citenamefont {Landaeta}, \citenamefont {Banda}, \citenamefont {Bannor},
  \citenamefont {Brando}, \citenamefont {Brydon}, \citenamefont {Hafner},
  \citenamefont {Küchler}, \citenamefont {Cardoso-Gil}, \citenamefont
  {Stockert}, \citenamefont {Mackenzie}, \citenamefont {Agterberg},
  \citenamefont {Geibel},\ and\ \citenamefont {Hassinger}}]{Khim2021}%
  \BibitemOpen
  \bibfield  {author} {\bibinfo {author} {\bibfnamefont {S.}~\bibnamefont
  {Khim}}, \bibinfo {author} {\bibfnamefont {J.~F.}\ \bibnamefont {Landaeta}},
  \bibinfo {author} {\bibfnamefont {J.}~\bibnamefont {Banda}}, \bibinfo
  {author} {\bibfnamefont {N.}~\bibnamefont {Bannor}}, \bibinfo {author}
  {\bibfnamefont {M.}~\bibnamefont {Brando}}, \bibinfo {author} {\bibfnamefont
  {P.~M.~R.}\ \bibnamefont {Brydon}}, \bibinfo {author} {\bibfnamefont
  {D.}~\bibnamefont {Hafner}}, \bibinfo {author} {\bibfnamefont
  {R.}~\bibnamefont {Küchler}}, \bibinfo {author} {\bibfnamefont
  {R.}~\bibnamefont {Cardoso-Gil}}, \bibinfo {author} {\bibfnamefont
  {U.}~\bibnamefont {Stockert}}, \bibinfo {author} {\bibfnamefont {A.~P.}\
  \bibnamefont {Mackenzie}}, \bibinfo {author} {\bibfnamefont {D.~F.}\
  \bibnamefont {Agterberg}}, \bibinfo {author} {\bibfnamefont {C.}~\bibnamefont
  {Geibel}},\ and\ \bibinfo {author} {\bibfnamefont {E.}~\bibnamefont
  {Hassinger}},\ }\bibfield  {title} {\bibinfo {title} {Field-induced
  transition within the superconducting state of {C}e{R}h$_2${A}s$_2$},\ }\href
  {https://doi.org/10.1126/science.abe7518} {\bibfield  {journal} {\bibinfo
  {journal} {Science}\ }\textbf {\bibinfo {volume} {373}},\ \bibinfo {pages}
  {1012} (\bibinfo {year} {2021})}\BibitemShut {NoStop}%
\bibitem [{\citenamefont {Schuberth}\ \emph {et~al.}(2016)\citenamefont
  {Schuberth}, \citenamefont {Tippmann}, \citenamefont {Steinke}, \citenamefont
  {Lausberg}, \citenamefont {Steppke}, \citenamefont {Brando}, \citenamefont
  {Krellner}, \citenamefont {Geibel}, \citenamefont {Yu}, \citenamefont {Si},\
  and\ \citenamefont {Steglich}}]{Schuberth2016}%
  \BibitemOpen
  \bibfield  {author} {\bibinfo {author} {\bibfnamefont {E.}~\bibnamefont
  {Schuberth}}, \bibinfo {author} {\bibfnamefont {M.}~\bibnamefont {Tippmann}},
  \bibinfo {author} {\bibfnamefont {L.}~\bibnamefont {Steinke}}, \bibinfo
  {author} {\bibfnamefont {S.}~\bibnamefont {Lausberg}}, \bibinfo {author}
  {\bibfnamefont {A.}~\bibnamefont {Steppke}}, \bibinfo {author} {\bibfnamefont
  {M.}~\bibnamefont {Brando}}, \bibinfo {author} {\bibfnamefont
  {C.}~\bibnamefont {Krellner}}, \bibinfo {author} {\bibfnamefont
  {C.}~\bibnamefont {Geibel}}, \bibinfo {author} {\bibfnamefont
  {R.}~\bibnamefont {Yu}}, \bibinfo {author} {\bibfnamefont {Q.}~\bibnamefont
  {Si}},\ and\ \bibinfo {author} {\bibfnamefont {F.}~\bibnamefont {Steglich}},\
  }\bibfield  {title} {\bibinfo {title} {Emergence of superconductivity in the
  canonical heavy-electron metal {YbRh}$_2${Si}$_2$},\ }\href@noop {}
  {\bibfield  {journal} {\bibinfo  {journal} {Science}\ }\textbf {\bibinfo
  {volume} {351}},\ \bibinfo {pages} {485} (\bibinfo {year}
  {2016})}\BibitemShut {NoStop}%
\bibitem [{\citenamefont {Mizushima}\ \emph {et~al.}(2016)\citenamefont
  {Mizushima}, \citenamefont {Tsutsumi}, \citenamefont {Kawakami},
  \citenamefont {Sato}, \citenamefont {Ichioka},\ and\ \citenamefont
  {Machida}}]{Mizushima2016}%
  \BibitemOpen
  \bibfield  {author} {\bibinfo {author} {\bibfnamefont {T.}~\bibnamefont
  {Mizushima}}, \bibinfo {author} {\bibfnamefont {Y.}~\bibnamefont {Tsutsumi}},
  \bibinfo {author} {\bibfnamefont {T.}~\bibnamefont {Kawakami}}, \bibinfo
  {author} {\bibfnamefont {M.}~\bibnamefont {Sato}}, \bibinfo {author}
  {\bibfnamefont {M.}~\bibnamefont {Ichioka}},\ and\ \bibinfo {author}
  {\bibfnamefont {K.}~\bibnamefont {Machida}},\ }\bibfield  {title} {\bibinfo
  {title} {Symmetry-protected topological superfluids and superconductors
  —{F}rom the {B}asics to {$^3$}{He}—},\ }\href@noop {} {\bibfield
  {journal} {\bibinfo  {journal} {Journal of the Physical Society of Japan}\
  }\textbf {\bibinfo {volume} {85}},\ \bibinfo {pages} {022001} (\bibinfo
  {year} {2016})}\BibitemShut {NoStop}%
\bibitem [{\citenamefont {M{\"a}kinen}\ \emph {et~al.}(2019)\citenamefont
  {M{\"a}kinen}, \citenamefont {Dmitriev}, \citenamefont {Nissinen},
  \citenamefont {Rysti}, \citenamefont {Volovik}, \citenamefont {Yudin},
  \citenamefont {Zhang},\ and\ \citenamefont {Eltsov}}]{Makinen2019}%
  \BibitemOpen
  \bibfield  {author} {\bibinfo {author} {\bibfnamefont {J.~T.}\ \bibnamefont
  {M{\"a}kinen}}, \bibinfo {author} {\bibfnamefont {V.~V.}\ \bibnamefont
  {Dmitriev}}, \bibinfo {author} {\bibfnamefont {J.}~\bibnamefont {Nissinen}},
  \bibinfo {author} {\bibfnamefont {J.}~\bibnamefont {Rysti}}, \bibinfo
  {author} {\bibfnamefont {G.~E.}\ \bibnamefont {Volovik}}, \bibinfo {author}
  {\bibfnamefont {A.~N.}\ \bibnamefont {Yudin}}, \bibinfo {author}
  {\bibfnamefont {K.}~\bibnamefont {Zhang}},\ and\ \bibinfo {author}
  {\bibfnamefont {V.~B.}\ \bibnamefont {Eltsov}},\ }\bibfield  {title}
  {\bibinfo {title} {Half-quantum vortices and walls bounded by strings in the
  polar-distorted phases of topological superfluid {$^3$}{He}},\ }\href@noop {}
  {\bibfield  {journal} {\bibinfo  {journal} {Nature Communications}\ }\textbf
  {\bibinfo {volume} {10}},\ \bibinfo {pages} {237} (\bibinfo {year}
  {2019})}\BibitemShut {NoStop}%
\bibitem [{\citenamefont {Volovik}(2021)}]{Volovik2021}%
  \BibitemOpen
  \bibfield  {author} {\bibinfo {author} {\bibfnamefont {G.~E.}\ \bibnamefont
  {Volovik}},\ }\bibfield  {title} {\bibinfo {title} {{$^3$}{H}e {U}niverse
  2020},\ }\href@noop {} {\bibfield  {journal} {\bibinfo  {journal} {Journal of
  Low Temperature Physics}\ }\textbf {\bibinfo {volume} {202}},\ \bibinfo
  {pages} {11} (\bibinfo {year} {2021})}\BibitemShut {NoStop}%
\bibitem [{\citenamefont {Heikkinen}\ \emph {et~al.}(2021)\citenamefont
  {Heikkinen}, \citenamefont {Casey}, \citenamefont {Levitin}, \citenamefont
  {Vorontsov}, \citenamefont {Sharma}, \citenamefont {Zhelev}, \citenamefont
  {Parpia},\ and\ \citenamefont {Saunders}}]{Heikkinen2021}%
  \BibitemOpen
  \bibfield  {author} {\bibinfo {author} {\bibfnamefont {P.~J.}\ \bibnamefont
  {Heikkinen}}, \bibinfo {author} {\bibfnamefont {A.}~\bibnamefont {Casey}},
  \bibinfo {author} {\bibfnamefont {X.}~\bibnamefont {Levitin}, \bibfnamefont
  {L.~V.and~Rojas}}, \bibinfo {author} {\bibfnamefont {A.}~\bibnamefont
  {Vorontsov}}, \bibinfo {author} {\bibfnamefont {P.}~\bibnamefont {Sharma}},
  \bibinfo {author} {\bibfnamefont {N.}~\bibnamefont {Zhelev}}, \bibinfo
  {author} {\bibfnamefont {J.~M.}\ \bibnamefont {Parpia}},\ and\ \bibinfo
  {author} {\bibfnamefont {J.}~\bibnamefont {Saunders}},\ }\bibfield  {title}
  {\bibinfo {title} {Fragility of surface states in topological superfluid
  {$^3$}{He}},\ }\href@noop {} {\bibfield  {journal} {\bibinfo  {journal}
  {Nature Communications}\ }\textbf {\bibinfo {volume} {12}},\ \bibinfo {pages}
  {1574} (\bibinfo {year} {2021})}\BibitemShut {NoStop}%
\bibitem [{\citenamefont {Nakatsuji}\ \emph {et~al.}(2008)\citenamefont
  {Nakatsuji}, \citenamefont {Kuga}, \citenamefont {Machida}, \citenamefont
  {Tayama}, \citenamefont {Sakakibara}, \citenamefont {Karaki}, \citenamefont
  {Ishimoto}, \citenamefont {Yonezawa}, \citenamefont {Maeno}, \citenamefont
  {Pearson}, \citenamefont {Lonzarich}, \citenamefont {Balicas}, \citenamefont
  {Lee},\ and\ \citenamefont {Fisk}}]{Nakatsuji2008}%
  \BibitemOpen
  \bibfield  {author} {\bibinfo {author} {\bibfnamefont {S.}~\bibnamefont
  {Nakatsuji}}, \bibinfo {author} {\bibfnamefont {K.}~\bibnamefont {Kuga}},
  \bibinfo {author} {\bibfnamefont {Y.}~\bibnamefont {Machida}}, \bibinfo
  {author} {\bibfnamefont {T.}~\bibnamefont {Tayama}}, \bibinfo {author}
  {\bibfnamefont {T.}~\bibnamefont {Sakakibara}}, \bibinfo {author}
  {\bibfnamefont {Y.}~\bibnamefont {Karaki}}, \bibinfo {author} {\bibfnamefont
  {H.}~\bibnamefont {Ishimoto}}, \bibinfo {author} {\bibfnamefont
  {S.}~\bibnamefont {Yonezawa}}, \bibinfo {author} {\bibfnamefont
  {Y.}~\bibnamefont {Maeno}}, \bibinfo {author} {\bibfnamefont
  {E.}~\bibnamefont {Pearson}}, \bibinfo {author} {\bibfnamefont {G.~G.}\
  \bibnamefont {Lonzarich}}, \bibinfo {author} {\bibfnamefont {L.}~\bibnamefont
  {Balicas}}, \bibinfo {author} {\bibfnamefont {H.}~\bibnamefont {Lee}},\ and\
  \bibinfo {author} {\bibfnamefont {Z.}~\bibnamefont {Fisk}},\ }\bibfield
  {title} {\bibinfo {title} {Superconductivity and quantum criticality in the
  heavy-fermion system $\beta${-YbAlB}$_4$},\ }\href@noop {} {\bibfield
  {journal} {\bibinfo  {journal} {Nature Physics}\ }\textbf {\bibinfo {volume}
  {4}},\ \bibinfo {pages} {603} (\bibinfo {year} {2008})}\BibitemShut {NoStop}%
\bibitem [{\citenamefont {Fuhrman}\ \emph {et~al.}(2021)\citenamefont
  {Fuhrman}, \citenamefont {Sidorenko}, \citenamefont {Hänel}, \citenamefont
  {Winkler}, \citenamefont {Prokofiev}, \citenamefont {Rodriguez-Rivera},
  \citenamefont {Qiu}, \citenamefont {Blaha}, \citenamefont {Si}, \citenamefont
  {Broholm},\ and\ \citenamefont {Paschen}}]{Fuhrman2021}%
  \BibitemOpen
  \bibfield  {author} {\bibinfo {author} {\bibfnamefont {W.~T.}\ \bibnamefont
  {Fuhrman}}, \bibinfo {author} {\bibfnamefont {A.}~\bibnamefont {Sidorenko}},
  \bibinfo {author} {\bibfnamefont {J.}~\bibnamefont {Hänel}}, \bibinfo
  {author} {\bibfnamefont {H.}~\bibnamefont {Winkler}}, \bibinfo {author}
  {\bibfnamefont {A.}~\bibnamefont {Prokofiev}}, \bibinfo {author}
  {\bibfnamefont {J.~A.}\ \bibnamefont {Rodriguez-Rivera}}, \bibinfo {author}
  {\bibfnamefont {Y.}~\bibnamefont {Qiu}}, \bibinfo {author} {\bibfnamefont
  {P.}~\bibnamefont {Blaha}}, \bibinfo {author} {\bibfnamefont
  {Q.}~\bibnamefont {Si}}, \bibinfo {author} {\bibfnamefont {C.~L.}\
  \bibnamefont {Broholm}},\ and\ \bibinfo {author} {\bibfnamefont
  {S.}~\bibnamefont {Paschen}},\ }\bibfield  {title} {\bibinfo {title}
  {Pristine quantum criticality in a {K}ondo semimetal},\ }\href@noop {}
  {\bibfield  {journal} {\bibinfo  {journal} {Science Advances}\ }\textbf
  {\bibinfo {volume} {7}},\ \bibinfo {pages} {eabf9134} (\bibinfo {year}
  {2021})}\BibitemShut {NoStop}%
\bibitem [{\citenamefont {Paschen}\ and\ \citenamefont
  {Si}(2021)}]{Paschen2021}%
  \BibitemOpen
  \bibfield  {author} {\bibinfo {author} {\bibfnamefont {S.}~\bibnamefont
  {Paschen}}\ and\ \bibinfo {author} {\bibfnamefont {Q.}~\bibnamefont {Si}},\
  }\bibfield  {title} {\bibinfo {title} {Quantum phases driven by strong
  correlations},\ }\href@noop {} {\bibfield  {journal} {\bibinfo  {journal}
  {Nature Reviews Physics}\ }\textbf {\bibinfo {volume} {3}},\ \bibinfo {pages}
  {9} (\bibinfo {year} {2021})}\BibitemShut {NoStop}%
\bibitem [{\citenamefont {Chekhovich}\ \emph {et~al.}(2013)\citenamefont
  {Chekhovich}, \citenamefont {Makhonin}, \citenamefont {Tartakovskii},
  \citenamefont {Yacoby}, \citenamefont {Bluhm}, \citenamefont {Nowack},\ and\
  \citenamefont {Vandersypen}}]{Chekhovich2013}%
  \BibitemOpen
  \bibfield  {author} {\bibinfo {author} {\bibfnamefont {E.~A.}\ \bibnamefont
  {Chekhovich}}, \bibinfo {author} {\bibfnamefont {M.~N.}\ \bibnamefont
  {Makhonin}}, \bibinfo {author} {\bibfnamefont {A.~I.}\ \bibnamefont
  {Tartakovskii}}, \bibinfo {author} {\bibfnamefont {A.}~\bibnamefont
  {Yacoby}}, \bibinfo {author} {\bibfnamefont {H.}~\bibnamefont {Bluhm}},
  \bibinfo {author} {\bibfnamefont {K.~C.}\ \bibnamefont {Nowack}},\ and\
  \bibinfo {author} {\bibfnamefont {L.~M.~K.}\ \bibnamefont {Vandersypen}},\
  }\bibfield  {title} {\bibinfo {title} {Nuclear spin effects in semiconductor
  quantum dots},\ }\href@noop {} {\bibfield  {journal} {\bibinfo  {journal}
  {Nature Materials}\ }\textbf {\bibinfo {volume} {12}},\ \bibinfo {pages}
  {494} (\bibinfo {year} {2013})}\BibitemShut {NoStop}%
\bibitem [{\citenamefont {Simon}\ \emph {et~al.}(2008)\citenamefont {Simon},
  \citenamefont {Braunecker},\ and\ \citenamefont {Loss}}]{Loss2018}%
  \BibitemOpen
  \bibfield  {author} {\bibinfo {author} {\bibfnamefont {P.}~\bibnamefont
  {Simon}}, \bibinfo {author} {\bibfnamefont {B.}~\bibnamefont {Braunecker}},\
  and\ \bibinfo {author} {\bibfnamefont {D.}~\bibnamefont {Loss}},\ }\bibfield
  {title} {\bibinfo {title} {Magnetic ordering of nuclear spins in an
  interacting two-dimensional electron gas},\ }\href@noop {} {\bibfield
  {journal} {\bibinfo  {journal} {Phys. Rev. B}\ }\textbf {\bibinfo {volume}
  {77}},\ \bibinfo {pages} {045108} (\bibinfo {year} {2008})}\BibitemShut
  {NoStop}%
\bibitem [{\citenamefont {Kikkawa}\ \emph {et~al.}(2021)\citenamefont
  {Kikkawa}, \citenamefont {Reitz}, \citenamefont {Ito}, \citenamefont
  {Makiuchi}, \citenamefont {Sugimoto}, \citenamefont {Tsunekawa},
  \citenamefont {Daimon}, \citenamefont {Oyanagi}, \citenamefont {Ramos},
  \citenamefont {Takahashi}, \citenamefont {Shiomi}, \citenamefont
  {Tserkovnyak},\ and\ \citenamefont {Saitoh}}]{Kikkawa2021}%
  \BibitemOpen
  \bibfield  {author} {\bibinfo {author} {\bibfnamefont {T.}~\bibnamefont
  {Kikkawa}}, \bibinfo {author} {\bibfnamefont {D.}~\bibnamefont {Reitz}},
  \bibinfo {author} {\bibfnamefont {H.}~\bibnamefont {Ito}}, \bibinfo {author}
  {\bibfnamefont {T.}~\bibnamefont {Makiuchi}}, \bibinfo {author}
  {\bibfnamefont {T.}~\bibnamefont {Sugimoto}}, \bibinfo {author}
  {\bibfnamefont {K.}~\bibnamefont {Tsunekawa}}, \bibinfo {author}
  {\bibfnamefont {S.}~\bibnamefont {Daimon}}, \bibinfo {author} {\bibfnamefont
  {K.}~\bibnamefont {Oyanagi}}, \bibinfo {author} {\bibfnamefont
  {R.}~\bibnamefont {Ramos}}, \bibinfo {author} {\bibfnamefont
  {S.}~\bibnamefont {Takahashi}}, \bibinfo {author} {\bibfnamefont
  {Y.}~\bibnamefont {Shiomi}}, \bibinfo {author} {\bibfnamefont
  {Y.}~\bibnamefont {Tserkovnyak}},\ and\ \bibinfo {author} {\bibfnamefont
  {E.}~\bibnamefont {Saitoh}},\ }\bibfield  {title} {\bibinfo {title}
  {Observation of nuclear-spin {S}eebeck effect},\ }\href@noop {} {\bibfield
  {journal} {\bibinfo  {journal} {Nature Communications}\ }\textbf {\bibinfo
  {volume} {12}},\ \bibinfo {pages} {4356} (\bibinfo {year}
  {2021})}\BibitemShut {NoStop}%
\bibitem [{\citenamefont {Eisenlohr}\ and\ \citenamefont
  {Vojta}(2021)}]{Eisenlohr2021}%
  \BibitemOpen
  \bibfield  {author} {\bibinfo {author} {\bibfnamefont {H.}~\bibnamefont
  {Eisenlohr}}\ and\ \bibinfo {author} {\bibfnamefont {M.}~\bibnamefont
  {Vojta}},\ }\bibfield  {title} {\bibinfo {title} {Limits to magnetic quantum
  criticality from nuclear spins},\ }\href
  {https://doi.org/10.1103/PhysRevB.103.064405} {\bibfield  {journal} {\bibinfo
   {journal} {Phys. Rev. B}\ }\textbf {\bibinfo {volume} {103}},\ \bibinfo
  {pages} {064405} (\bibinfo {year} {2021})}\BibitemShut {NoStop}%
\bibitem [{\citenamefont {Muhonen}\ \emph {et~al.}(2012)\citenamefont
  {Muhonen}, \citenamefont {Meschke},\ and\ \citenamefont
  {Pekola}}]{Muhonen_2012}%
  \BibitemOpen
  \bibfield  {author} {\bibinfo {author} {\bibfnamefont {J.~T.}\ \bibnamefont
  {Muhonen}}, \bibinfo {author} {\bibfnamefont {M.}~\bibnamefont {Meschke}},\
  and\ \bibinfo {author} {\bibfnamefont {J.~P.}\ \bibnamefont {Pekola}},\
  }\bibfield  {title} {\bibinfo {title} {Micrometre-scale refrigerators},\
  }\href@noop {} {\bibfield  {journal} {\bibinfo  {journal} {Reports on
  Progress in Physics}\ }\textbf {\bibinfo {volume} {75}},\ \bibinfo {pages}
  {046501} (\bibinfo {year} {2012})}\BibitemShut {NoStop}%
\bibitem [{\citenamefont {Pekola}\ and\ \citenamefont
  {Karimi}(2021)}]{Pekola2021}%
  \BibitemOpen
  \bibfield  {author} {\bibinfo {author} {\bibfnamefont {J.~P.}\ \bibnamefont
  {Pekola}}\ and\ \bibinfo {author} {\bibfnamefont {B.}~\bibnamefont
  {Karimi}},\ }\bibfield  {title} {\bibinfo {title} {Colloquium: {Q}uantum heat
  transport in condensed matter systems},\ }\href@noop {} {\bibfield  {journal}
  {\bibinfo  {journal} {Rev. Mod. Phys.}\ }\textbf {\bibinfo {volume} {93}},\
  \bibinfo {pages} {041001} (\bibinfo {year} {2021})}\BibitemShut {NoStop}%
\bibitem [{\citenamefont {Clark}\ \emph {et~al.}(2010)\citenamefont {Clark},
  \citenamefont {Schwarzwälder}, \citenamefont {Bandi}, \citenamefont
  {Maradan},\ and\ \citenamefont {Zumb{\"u}hl}}]{Clark2010}%
  \BibitemOpen
  \bibfield  {author} {\bibinfo {author} {\bibfnamefont {A.~C.}\ \bibnamefont
  {Clark}}, \bibinfo {author} {\bibfnamefont {K.~K.}\ \bibnamefont
  {Schwarzwälder}}, \bibinfo {author} {\bibfnamefont {T.}~\bibnamefont
  {Bandi}}, \bibinfo {author} {\bibfnamefont {D.}~\bibnamefont {Maradan}},\
  and\ \bibinfo {author} {\bibfnamefont {D.~M.}\ \bibnamefont {Zumb{\"u}hl}},\
  }\bibfield  {title} {\bibinfo {title} {Method for cooling nanostructures to
  microkelvin temperatures},\ }\href@noop {} {\bibfield  {journal} {\bibinfo
  {journal} {Review of Scientific Instruments}\ }\textbf {\bibinfo {volume}
  {81}},\ \bibinfo {pages} {103904} (\bibinfo {year} {2010})}\BibitemShut
  {NoStop}%
\bibitem [{\citenamefont {Palma}\ \emph {et~al.}(2017)\citenamefont {Palma},
  \citenamefont {Maradan}, \citenamefont {Casparis}, \citenamefont {Liu},
  \citenamefont {Froning},\ and\ \citenamefont {Zumbühl}}]{Palma2017}%
  \BibitemOpen
  \bibfield  {author} {\bibinfo {author} {\bibfnamefont {M.}~\bibnamefont
  {Palma}}, \bibinfo {author} {\bibfnamefont {D.}~\bibnamefont {Maradan}},
  \bibinfo {author} {\bibfnamefont {L.}~\bibnamefont {Casparis}}, \bibinfo
  {author} {\bibfnamefont {T.-M.}\ \bibnamefont {Liu}}, \bibinfo {author}
  {\bibfnamefont {F.~N.~M.}\ \bibnamefont {Froning}},\ and\ \bibinfo {author}
  {\bibfnamefont {D.~M.}\ \bibnamefont {Zumbühl}},\ }\bibfield  {title}
  {\bibinfo {title} {Magnetic cooling for microkelvin nanoelectronics on a
  cryofree platform},\ }\href@noop {} {\bibfield  {journal} {\bibinfo
  {journal} {Review of Scientific Instruments}\ }\textbf {\bibinfo {volume}
  {88}},\ \bibinfo {pages} {043902} (\bibinfo {year} {2017})}\BibitemShut
  {NoStop}%
\bibitem [{\citenamefont {Jones}\ \emph {et~al.}(2020)\citenamefont {Jones},
  \citenamefont {Scheller}, \citenamefont {Prance}, \citenamefont {Kalyoncu},
  \citenamefont {Zumb{\"u}hl},\ and\ \citenamefont {Haley}}]{Jones2020-ie}%
  \BibitemOpen
  \bibfield  {author} {\bibinfo {author} {\bibfnamefont {A.~T.}\ \bibnamefont
  {Jones}}, \bibinfo {author} {\bibfnamefont {C.~P.}\ \bibnamefont {Scheller}},
  \bibinfo {author} {\bibfnamefont {J.~R.}\ \bibnamefont {Prance}}, \bibinfo
  {author} {\bibfnamefont {Y.~B.}\ \bibnamefont {Kalyoncu}}, \bibinfo {author}
  {\bibfnamefont {D.~M.}\ \bibnamefont {Zumb{\"u}hl}},\ and\ \bibinfo {author}
  {\bibfnamefont {R.~P.}\ \bibnamefont {Haley}},\ }\bibfield  {title} {\bibinfo
  {title} {Progress in cooling nanoelectronic devices to {Ultra-Low}
  temperatures},\ }\href@noop {} {\bibfield  {journal} {\bibinfo  {journal}
  {Journal of Low Temperature Physics}\ }\textbf {\bibinfo {volume} {201}},\
  \bibinfo {pages} {772} (\bibinfo {year} {2020})}\BibitemShut {NoStop}%
\bibitem [{\citenamefont {Sarsby}\ \emph {et~al.}(2020)\citenamefont {Sarsby},
  \citenamefont {Yurttag{\"u}l},\ and\ \citenamefont {Geresdi}}]{Sarsby2020}%
  \BibitemOpen
  \bibfield  {author} {\bibinfo {author} {\bibfnamefont {M.}~\bibnamefont
  {Sarsby}}, \bibinfo {author} {\bibfnamefont {N.}~\bibnamefont
  {Yurttag{\"u}l}},\ and\ \bibinfo {author} {\bibfnamefont {A.}~\bibnamefont
  {Geresdi}},\ }\bibfield  {title} {\bibinfo {title} {500 microkelvin
  nanoelectronics},\ }\href@noop {} {\bibfield  {journal} {\bibinfo  {journal}
  {Nature Communications}\ }\textbf {\bibinfo {volume} {11}},\ \bibinfo {pages}
  {1492} (\bibinfo {year} {2020})}\BibitemShut {NoStop}%
\bibitem [{\citenamefont {Levitin}\ \emph {et~al.}(2022)\citenamefont
  {Levitin}, \citenamefont {van~der Vliet}, \citenamefont {Theisen},
  \citenamefont {Dimitriadis}, \citenamefont {Lucas}, \citenamefont {Corcoles},
  \citenamefont {Ny{\'e}ki}, \citenamefont {Casey}, \citenamefont {Creeth},
  \citenamefont {Farrer}, \citenamefont {Ritchie}, \citenamefont {Nicholls},\
  and\ \citenamefont {Saunders}}]{Levitin2022}%
  \BibitemOpen
  \bibfield  {author} {\bibinfo {author} {\bibfnamefont {L.~V.}\ \bibnamefont
  {Levitin}}, \bibinfo {author} {\bibfnamefont {H.}~\bibnamefont {van~der
  Vliet}}, \bibinfo {author} {\bibfnamefont {T.}~\bibnamefont {Theisen}},
  \bibinfo {author} {\bibfnamefont {S.}~\bibnamefont {Dimitriadis}}, \bibinfo
  {author} {\bibfnamefont {M.}~\bibnamefont {Lucas}}, \bibinfo {author}
  {\bibfnamefont {A.~D.}\ \bibnamefont {Corcoles}}, \bibinfo {author}
  {\bibfnamefont {J.}~\bibnamefont {Ny{\'e}ki}}, \bibinfo {author}
  {\bibfnamefont {A.~J.}\ \bibnamefont {Casey}}, \bibinfo {author}
  {\bibfnamefont {G.}~\bibnamefont {Creeth}}, \bibinfo {author} {\bibfnamefont
  {I.}~\bibnamefont {Farrer}}, \bibinfo {author} {\bibfnamefont {D.~A.}\
  \bibnamefont {Ritchie}}, \bibinfo {author} {\bibfnamefont {J.~T.}\
  \bibnamefont {Nicholls}},\ and\ \bibinfo {author} {\bibfnamefont
  {J.}~\bibnamefont {Saunders}},\ }\bibfield  {title} {\bibinfo {title}
  {Cooling low-dimensional electron systems into the microkelvin regime},\
  }\href@noop {} {\bibfield  {journal} {\bibinfo  {journal} {Nature
  Communications}\ }\textbf {\bibinfo {volume} {13}},\ \bibinfo {pages} {667}
  (\bibinfo {year} {2022})}\BibitemShut {NoStop}%
\bibitem [{\citenamefont {Knapp}\ \emph {et~al.}()\citenamefont {Knapp} \emph
  {et~al.}}]{Knapp2022}%
  \BibitemOpen
  \bibfield  {author} {\bibinfo {author} {\bibfnamefont {J.}~\bibnamefont
  {Knapp}} \emph {et~al.},\ }\bibinfo {title} {(unpublished)}\BibitemShut
  {NoStop}%
\bibitem [{\citenamefont {Bangma}\ \emph {et~al.}()\citenamefont {Bangma} \emph
  {et~al.}}]{Bangma2022}%
  \BibitemOpen
\bibfield  {title} {  }\bibfield  {author} {\bibinfo {author} {\bibfnamefont
  {F.}~\bibnamefont {Bangma}} \emph {et~al.},\ }\bibinfo {title}
  {(unpublished)}\BibitemShut {NoStop}%
\bibitem [{\citenamefont {Batey}\ \emph {et~al.}(2013)\citenamefont {Batey},
  \citenamefont {Casey}, \citenamefont {Cuthbert}, \citenamefont {Matthews},
  \citenamefont {Saunders},\ and\ \citenamefont {Shibahara}}]{Batey_2013}%
  \BibitemOpen
\bibfield  {title} {  }\bibfield  {author} {\bibinfo {author} {\bibfnamefont
  {G.}~\bibnamefont {Batey}}, \bibinfo {author} {\bibfnamefont
  {A.}~\bibnamefont {Casey}}, \bibinfo {author} {\bibfnamefont {M.~N.}\
  \bibnamefont {Cuthbert}}, \bibinfo {author} {\bibfnamefont {A.~J.}\
  \bibnamefont {Matthews}}, \bibinfo {author} {\bibfnamefont {J.}~\bibnamefont
  {Saunders}},\ and\ \bibinfo {author} {\bibfnamefont {A.}~\bibnamefont
  {Shibahara}},\ }\bibfield  {title} {\bibinfo {title} {A microkelvin
  cryogen-free experimental platform with integrated noise thermometry},\
  }\href@noop {} {\bibfield  {journal} {\bibinfo  {journal} {New Journal of
  Physics}\ }\textbf {\bibinfo {volume} {15}},\ \bibinfo {pages} {113034}
  (\bibinfo {year} {2013})}\BibitemShut {NoStop}%
\bibitem [{\citenamefont {Parpia}\ \emph {et~al.}(1985)\citenamefont {Parpia},
  \citenamefont {Kirk}, \citenamefont {Kobiela}, \citenamefont {Rhodes},
  \citenamefont {Olejniczak},\ and\ \citenamefont {Parker}}]{Parpia1985}%
  \BibitemOpen
  \bibfield  {author} {\bibinfo {author} {\bibfnamefont {J.~M.}\ \bibnamefont
  {Parpia}}, \bibinfo {author} {\bibfnamefont {W.~P.}\ \bibnamefont {Kirk}},
  \bibinfo {author} {\bibfnamefont {P.~S.}\ \bibnamefont {Kobiela}}, \bibinfo
  {author} {\bibfnamefont {T.~L.}\ \bibnamefont {Rhodes}}, \bibinfo {author}
  {\bibfnamefont {Z.}~\bibnamefont {Olejniczak}},\ and\ \bibinfo {author}
  {\bibfnamefont {G.~N.}\ \bibnamefont {Parker}},\ }\bibfield  {title}
  {\bibinfo {title} {Optimization procedure for the cooling of liquid 3{H}e by
  adiabatic demagnetization of praseodymium nickel},\ }\href@noop {} {\bibfield
   {journal} {\bibinfo  {journal} {Review of Scientific Instruments}\ }\textbf
  {\bibinfo {volume} {56}},\ \bibinfo {pages} {437} (\bibinfo {year}
  {1985})}\BibitemShut {NoStop}%
\bibitem [{\citenamefont {Schmoranzer}\ \emph {et~al.}(2019)\citenamefont
  {Schmoranzer}, \citenamefont {Luck}, \citenamefont {Collin},\ and\
  \citenamefont {Fefferman}}]{SCHMORANZER2019102}%
  \BibitemOpen
  \bibfield  {author} {\bibinfo {author} {\bibfnamefont {D.}~\bibnamefont
  {Schmoranzer}}, \bibinfo {author} {\bibfnamefont {A.}~\bibnamefont {Luck}},
  \bibinfo {author} {\bibfnamefont {E.}~\bibnamefont {Collin}},\ and\ \bibinfo
  {author} {\bibfnamefont {A.}~\bibnamefont {Fefferman}},\ }\bibfield  {title}
  {\bibinfo {title} {Cryogenic broadband vibration measurement on a
  cryogen-free dilution refrigerator},\ }\href@noop {} {\bibfield  {journal}
  {\bibinfo  {journal} {Cryogenics}\ }\textbf {\bibinfo {volume} {98}},\
  \bibinfo {pages} {102} (\bibinfo {year} {2019})}\BibitemShut {NoStop}%
\bibitem [{\citenamefont {Todoshchenko}\ \emph {et~al.}(2014)\citenamefont
  {Todoshchenko}, \citenamefont {Kaikkonen}, \citenamefont {Blaauwgeers},
  \citenamefont {Hakonen},\ and\ \citenamefont {Savin}}]{Todoshchenko2014}%
  \BibitemOpen
  \bibfield  {author} {\bibinfo {author} {\bibfnamefont {I.}~\bibnamefont
  {Todoshchenko}}, \bibinfo {author} {\bibfnamefont {J.-P.}\ \bibnamefont
  {Kaikkonen}}, \bibinfo {author} {\bibfnamefont {R.}~\bibnamefont
  {Blaauwgeers}}, \bibinfo {author} {\bibfnamefont {P.~J.}\ \bibnamefont
  {Hakonen}},\ and\ \bibinfo {author} {\bibfnamefont {A.}~\bibnamefont
  {Savin}},\ }\bibfield  {title} {\bibinfo {title} {Dry demagnetization
  cryostat for sub-millikelvin helium experiments: {R}efrigeration and
  thermometry},\ }\href@noop {} {\bibfield  {journal} {\bibinfo  {journal}
  {Review of Scientific Instruments}\ }\textbf {\bibinfo {volume} {85}},\
  \bibinfo {pages} {085106} (\bibinfo {year} {2014})}\BibitemShut {NoStop}%
\bibitem [{\citenamefont {Yan}\ \emph {et~al.}(2021)\citenamefont {Yan},
  \citenamefont {Yao}, \citenamefont {Shvarts}, \citenamefont {Du},\ and\
  \citenamefont {Lin}}]{Yan2021}%
  \BibitemOpen
  \bibfield  {author} {\bibinfo {author} {\bibfnamefont {J.}~\bibnamefont
  {Yan}}, \bibinfo {author} {\bibfnamefont {J.}~\bibnamefont {Yao}}, \bibinfo
  {author} {\bibfnamefont {V.}~\bibnamefont {Shvarts}}, \bibinfo {author}
  {\bibfnamefont {R.-R.}\ \bibnamefont {Du}},\ and\ \bibinfo {author}
  {\bibfnamefont {X.}~\bibnamefont {Lin}},\ }\bibfield  {title} {\bibinfo
  {title} {Cryogen-free one hundred microkelvin refrigerator},\ }\href@noop {}
  {\bibfield  {journal} {\bibinfo  {journal} {Review of Scientific
  Instruments}\ }\textbf {\bibinfo {volume} {92}},\ \bibinfo {pages} {025120}
  (\bibinfo {year} {2021})}\BibitemShut {NoStop}%
\bibitem [{\citenamefont {Toda}\ \emph {et~al.}(2018)\citenamefont {Toda},
  \citenamefont {Murakawa},\ and\ \citenamefont {Fukuyama}}]{Toda_2018}%
  \BibitemOpen
  \bibfield  {author} {\bibinfo {author} {\bibfnamefont {R.}~\bibnamefont
  {Toda}}, \bibinfo {author} {\bibfnamefont {S.}~\bibnamefont {Murakawa}},\
  and\ \bibinfo {author} {\bibfnamefont {H.}~\bibnamefont {Fukuyama}},\
  }\bibfield  {title} {\bibinfo {title} {Design and expected performance of a
  compact and continuous nuclear demagnetization refrigerator for sub-{mK}
  applications},\ }\href@noop {} {\bibfield  {journal} {\bibinfo  {journal}
  {Journal of Physics: Conference Series}\ }\textbf {\bibinfo {volume} {969}},\
  \bibinfo {pages} {012093} (\bibinfo {year} {2018})}\BibitemShut {NoStop}%
\bibitem [{\citenamefont {Schmoranzer}\ \emph {et~al.}(2020)\citenamefont
  {Schmoranzer}, \citenamefont {Butterworth}, \citenamefont {Triqueneaux},
  \citenamefont {Collin},\ and\ \citenamefont {Fefferman}}]{SCHMORANZER2020}%
  \BibitemOpen
  \bibfield  {author} {\bibinfo {author} {\bibfnamefont {D.}~\bibnamefont
  {Schmoranzer}}, \bibinfo {author} {\bibfnamefont {J.}~\bibnamefont
  {Butterworth}}, \bibinfo {author} {\bibfnamefont {S.}~\bibnamefont
  {Triqueneaux}}, \bibinfo {author} {\bibfnamefont {E.}~\bibnamefont
  {Collin}},\ and\ \bibinfo {author} {\bibfnamefont {A.}~\bibnamefont
  {Fefferman}},\ }\bibfield  {title} {\bibinfo {title} {Design evaluation of
  serial and parallel sub-m{K} continuous nuclear demagnetization
  refrigerators},\ }\href@noop {} {\bibfield  {journal} {\bibinfo  {journal}
  {Cryogenics}\ }\textbf {\bibinfo {volume} {110}},\ \bibinfo {pages} {103119}
  (\bibinfo {year} {2020})}\BibitemShut {NoStop}%
\bibitem [{\citenamefont {Andres}\ and\ \citenamefont
  {Darack}(1977)}]{Andres1977}%
  \BibitemOpen
  \bibfield  {author} {\bibinfo {author} {\bibfnamefont {K.}~\bibnamefont
  {Andres}}\ and\ \bibinfo {author} {\bibfnamefont {S.}~\bibnamefont
  {Darack}},\ }\bibfield  {title} {\bibinfo {title} {Cooling of $^3${He} to 1
  m{K} by nuclear demagnetization of {PrNi$_5$}},\ }\href@noop {} {\bibfield
  {journal} {\bibinfo  {journal} {Physica B+C}\ }\textbf {\bibinfo {volume}
  {86-88}},\ \bibinfo {pages} {1071} (\bibinfo {year} {1977})}\BibitemShut
  {NoStop}%
\bibitem [{\citenamefont {Mueller}\ \emph {et~al.}(1980)\citenamefont
  {Mueller}, \citenamefont {Buchal}, \citenamefont {Folle}, \citenamefont
  {Kubota},\ and\ \citenamefont {Pobell}}]{MUELLER1980}%
  \BibitemOpen
  \bibfield  {author} {\bibinfo {author} {\bibfnamefont {R.}~\bibnamefont
  {Mueller}}, \bibinfo {author} {\bibfnamefont {C.}~\bibnamefont {Buchal}},
  \bibinfo {author} {\bibfnamefont {H.}~\bibnamefont {Folle}}, \bibinfo
  {author} {\bibfnamefont {M.}~\bibnamefont {Kubota}},\ and\ \bibinfo {author}
  {\bibfnamefont {F.}~\bibnamefont {Pobell}},\ }\bibfield  {title} {\bibinfo
  {title} {A double-stage nuclear demagnetization refrigerator},\ }\href@noop
  {} {\bibfield  {journal} {\bibinfo  {journal} {Cryogenics}\ }\textbf
  {\bibinfo {volume} {20}},\ \bibinfo {pages} {395} (\bibinfo {year}
  {1980})}\BibitemShut {NoStop}%
\bibitem [{\citenamefont {Folle}\ \emph {et~al.}(1981)\citenamefont {Folle},
  \citenamefont {Kubota}, \citenamefont {Buchal}, \citenamefont {Mueller},\
  and\ \citenamefont {Pobell}}]{Folle1981}%
  \BibitemOpen
  \bibfield  {author} {\bibinfo {author} {\bibfnamefont {H.~R.}\ \bibnamefont
  {Folle}}, \bibinfo {author} {\bibfnamefont {M.}~\bibnamefont {Kubota}},
  \bibinfo {author} {\bibfnamefont {C.}~\bibnamefont {Buchal}}, \bibinfo
  {author} {\bibfnamefont {R.~M.}\ \bibnamefont {Mueller}},\ and\ \bibinfo
  {author} {\bibfnamefont {F.}~\bibnamefont {Pobell}},\ }\bibfield  {title}
  {\bibinfo {title} {Nuclear refrigeration properties of {PrNi$_5$}},\
  }\href@noop {} {\bibfield  {journal} {\bibinfo  {journal} {Zeitschrift
  f{\"u}r Physik B Condensed Matter}\ }\textbf {\bibinfo {volume} {41}},\
  \bibinfo {pages} {223} (\bibinfo {year} {1981})}\BibitemShut {NoStop}%
\bibitem [{\citenamefont {Greywall}(1985)}]{Greywall1985}%
  \BibitemOpen
  \bibfield  {author} {\bibinfo {author} {\bibfnamefont {D.~S.}\ \bibnamefont
  {Greywall}},\ }\bibfield  {title} {\bibinfo {title} {$^{3}${He} melting-curve
  thermometry at millikelvin temperatures},\ }\href@noop {} {\bibfield
  {journal} {\bibinfo  {journal} {Phys. Rev. B}\ }\textbf {\bibinfo {volume}
  {31}},\ \bibinfo {pages} {2675} (\bibinfo {year} {1985})}\BibitemShut
  {NoStop}%
\bibitem [{\citenamefont {Wiegers}\ \emph {et~al.}(1990)\citenamefont
  {Wiegers}, \citenamefont {Hata}, \citenamefont {Kranenburg}, \citenamefont
  {{van de Haar}}, \citenamefont {Jochemsen},\ and\ \citenamefont
  {Frossati}}]{WIEGERS1990}%
  \BibitemOpen
  \bibfield  {author} {\bibinfo {author} {\bibfnamefont {S.}~\bibnamefont
  {Wiegers}}, \bibinfo {author} {\bibfnamefont {T.}~\bibnamefont {Hata}},
  \bibinfo {author} {\bibfnamefont {C.}~\bibnamefont {Kranenburg}}, \bibinfo
  {author} {\bibfnamefont {P.}~\bibnamefont {{van de Haar}}}, \bibinfo {author}
  {\bibfnamefont {R.}~\bibnamefont {Jochemsen}},\ and\ \bibinfo {author}
  {\bibfnamefont {G.}~\bibnamefont {Frossati}},\ }\bibfield  {title} {\bibinfo
  {title} {Compact {PrNi$_5$} nuclear demagnetization cryostat},\ }\href@noop
  {} {\bibfield  {journal} {\bibinfo  {journal} {Cryogenics}\ }\textbf
  {\bibinfo {volume} {30}},\ \bibinfo {pages} {770} (\bibinfo {year}
  {1990})}\BibitemShut {NoStop}%
\bibitem [{\citenamefont {Lusher}\ \emph {et~al.}(2000)\citenamefont {Lusher},
  \citenamefont {Li}, \citenamefont {Maidanov}, \citenamefont {Digby},
  \citenamefont {Dyball}, \citenamefont {Casey}, \citenamefont {Ny{\'{e}}ki},
  \citenamefont {Dmitriev}, \citenamefont {Cowan},\ and\ \citenamefont
  {Saunders}}]{Lusher2000}%
  \BibitemOpen
  \bibfield  {author} {\bibinfo {author} {\bibfnamefont {C.~P.}\ \bibnamefont
  {Lusher}}, \bibinfo {author} {\bibfnamefont {J.}~\bibnamefont {Li}}, \bibinfo
  {author} {\bibfnamefont {V.~A.}\ \bibnamefont {Maidanov}}, \bibinfo {author}
  {\bibfnamefont {M.~E.}\ \bibnamefont {Digby}}, \bibinfo {author}
  {\bibfnamefont {H.}~\bibnamefont {Dyball}}, \bibinfo {author} {\bibfnamefont
  {A.}~\bibnamefont {Casey}}, \bibinfo {author} {\bibfnamefont
  {J.}~\bibnamefont {Ny{\'{e}}ki}}, \bibinfo {author} {\bibfnamefont {V.~V.}\
  \bibnamefont {Dmitriev}}, \bibinfo {author} {\bibfnamefont {B.~P.}\
  \bibnamefont {Cowan}},\ and\ \bibinfo {author} {\bibfnamefont
  {J.}~\bibnamefont {Saunders}},\ }\bibfield  {title} {\bibinfo {title}
  {Current sensing noise thermometry using a low {Tc} {DC} {SQUID}
  preamplifier},\ }\href {https://doi.org/10.1088/0957-0233/12/1/301}
  {\bibfield  {journal} {\bibinfo  {journal} {Measurement Science and
  Technology}\ }\textbf {\bibinfo {volume} {12}},\ \bibinfo {pages} {1}
  (\bibinfo {year} {2000})}\BibitemShut {NoStop}%
\bibitem [{\citenamefont {Shibahara}\ \emph {et~al.}(2016)\citenamefont
  {Shibahara}, \citenamefont {Hahtela}, \citenamefont {Engert}, \citenamefont
  {van~der Vliet}, \citenamefont {Levitin}, \citenamefont {Casey},
  \citenamefont {Lusher}, \citenamefont {Saunders}, \citenamefont {Drung},\
  and\ \citenamefont {Schurig}}]{Shibahara2016}%
  \BibitemOpen
  \bibfield  {author} {\bibinfo {author} {\bibfnamefont {A.}~\bibnamefont
  {Shibahara}}, \bibinfo {author} {\bibfnamefont {O.}~\bibnamefont {Hahtela}},
  \bibinfo {author} {\bibfnamefont {J.}~\bibnamefont {Engert}}, \bibinfo
  {author} {\bibfnamefont {H.}~\bibnamefont {van~der Vliet}}, \bibinfo {author}
  {\bibfnamefont {L.~V.}\ \bibnamefont {Levitin}}, \bibinfo {author}
  {\bibfnamefont {A.}~\bibnamefont {Casey}}, \bibinfo {author} {\bibfnamefont
  {C.~P.}\ \bibnamefont {Lusher}}, \bibinfo {author} {\bibfnamefont
  {J.}~\bibnamefont {Saunders}}, \bibinfo {author} {\bibfnamefont
  {D.}~\bibnamefont {Drung}},\ and\ \bibinfo {author} {\bibfnamefont
  {T.}~\bibnamefont {Schurig}},\ }\bibfield  {title} {\bibinfo {title} {Primary
  current-sensing noise thermometry in the millikelvin regime},\ }\href@noop {}
  {\bibfield  {journal} {\bibinfo  {journal} {Philosophical Transactions of the
  Royal Society A: Mathematical, Physical and Engineering Sciences}\ }\textbf
  {\bibinfo {volume} {374}},\ \bibinfo {pages} {20150054} (\bibinfo {year}
  {2016})}\BibitemShut {NoStop}%
\bibitem [{Note2()}]{Note2}%
  \BibitemOpen
  \bibinfo {note} {Oxford Instruments NanoScience, UK.
  https://www.oxinst.com/}\BibitemShut {NoStop}%
\bibitem [{\citenamefont {Rybalko}\ and\ \citenamefont
  {Sterin}(1996)}]{Rybalko1996}%
  \BibitemOpen
  \bibfield  {author} {\bibinfo {author} {\bibfnamefont {A.~S.}\ \bibnamefont
  {Rybalko}}\ and\ \bibinfo {author} {\bibfnamefont {M.~B.}\ \bibnamefont
  {Sterin}},\ }\bibfield  {title} {\bibinfo {title} {Electrical and thermal
  conductivity of the conical thermal contacts used at ultra low temperature},\
  }\href@noop {} {\bibfield  {journal} {\bibinfo  {journal} {Fizika Nizkikh
  Temperatur}\ }\textbf {\bibinfo {volume} {22}},\ \bibinfo {pages} {1095}
  (\bibinfo {year} {1996})}\BibitemShut {NoStop}%
\bibitem [{Note3()}]{Note3}%
  \BibitemOpen
  \bibinfo {note} {Luvata Wolverhampton Ltd., UK.\protect \newline
  https://www.luvata.com/}\BibitemShut {NoStop}%
\bibitem [{Note4()}]{Note4}%
  \BibitemOpen
  \bibinfo {note} {Twickenham Plating Group, UK.\protect \newline
  http://twickenham.co.uk/}\BibitemShut {NoStop}%
\bibitem [{Note5()}]{Note5}%
  \BibitemOpen
  \bibinfo {note} {AMES Laboratory, U.S. Department of Energy, Iowa State
  University}\BibitemShut {NoStop}%
\bibitem [{\citenamefont {Kubota}\ \emph {et~al.}(1980)\citenamefont {Kubota},
  \citenamefont {Folle}, \citenamefont {Buchal}, \citenamefont {Mueller},\ and\
  \citenamefont {Pobell}}]{Kubota1980}%
  \BibitemOpen
  \bibfield  {author} {\bibinfo {author} {\bibfnamefont {M.}~\bibnamefont
  {Kubota}}, \bibinfo {author} {\bibfnamefont {H.~R.}\ \bibnamefont {Folle}},
  \bibinfo {author} {\bibfnamefont {C.}~\bibnamefont {Buchal}}, \bibinfo
  {author} {\bibfnamefont {R.~M.}\ \bibnamefont {Mueller}},\ and\ \bibinfo
  {author} {\bibfnamefont {F.}~\bibnamefont {Pobell}},\ }\bibfield  {title}
  {\bibinfo {title} {Nuclear magnetic ordering in {PrNi}$_5$ at 0.4 m{K}},\
  }\href@noop {} {\bibfield  {journal} {\bibinfo  {journal} {Phys. Rev. Lett.}\
  }\textbf {\bibinfo {volume} {45}},\ \bibinfo {pages} {1812} (\bibinfo {year}
  {1980})}\BibitemShut {NoStop}%
\bibitem [{\citenamefont {Korringa}(1950)}]{KORRINGA1950}%
  \BibitemOpen
  \bibfield  {author} {\bibinfo {author} {\bibfnamefont {J.}~\bibnamefont
  {Korringa}},\ }\bibfield  {title} {\bibinfo {title} {Nuclear magnetic
  relaxation and resonance line shift in metals},\ }\href@noop {} {\bibfield
  {journal} {\bibinfo  {journal} {Physica}\ }\textbf {\bibinfo {volume} {16}},\
  \bibinfo {pages} {601} (\bibinfo {year} {1950})}\BibitemShut {NoStop}%
\bibitem [{\citenamefont {Herrmannsd{\"o}rfer}\ \emph
  {et~al.}(1994)\citenamefont {Herrmannsd{\"o}rfer}, \citenamefont {Uniewski},\
  and\ \citenamefont {Pobell}}]{Herrmannsdorfer1994}%
  \BibitemOpen
  \bibfield  {author} {\bibinfo {author} {\bibfnamefont {T.}~\bibnamefont
  {Herrmannsd{\"o}rfer}}, \bibinfo {author} {\bibfnamefont {H.}~\bibnamefont
  {Uniewski}},\ and\ \bibinfo {author} {\bibfnamefont {F.}~\bibnamefont
  {Pobell}},\ }\bibfield  {title} {\bibinfo {title} {Nuclear ferromagnetic
  ordering of $^{141}${Pr} in the diluted {V}an {V}leck paramagnets
  {Pr}$_{1-x}${Y}$_x${Ni}$_5$},\ }\href@noop {} {\bibfield  {journal} {\bibinfo
   {journal} {Journal of Low Temperature Physics}\ }\textbf {\bibinfo {volume}
  {97}},\ \bibinfo {pages} {189} (\bibinfo {year} {1994})}\BibitemShut
  {NoStop}%
\bibitem [{\citenamefont {Takimoto}\ \emph {et~al.}(2022)\citenamefont
  {Takimoto}, \citenamefont {Toda}, \citenamefont {Murakawa},\ and\
  \citenamefont {Fukuyama}}]{Takimoto2022}%
  \BibitemOpen
  \bibfield  {author} {\bibinfo {author} {\bibfnamefont {S.}~\bibnamefont
  {Takimoto}}, \bibinfo {author} {\bibfnamefont {R.}~\bibnamefont {Toda}},
  \bibinfo {author} {\bibfnamefont {S.}~\bibnamefont {Murakawa}},\ and\
  \bibinfo {author} {\bibfnamefont {H.}~\bibnamefont {Fukuyama}},\ }\bibfield
  {title} {\bibinfo {title} {Construction of continuous magnetic cooling
  apparatus with {Z}inc soldered {PrNi$_5$} nuclear stages},\ }\href
  {https://doi.org/10.1007/s10909-022-02801-0} {\bibfield  {journal} {\bibinfo
  {journal} {Journal of Low Temperature Physics}\ }\textbf {\bibinfo {volume}
  {208}},\ \bibinfo {pages} {492} (\bibinfo {year} {2022})}\BibitemShut
  {NoStop}%
\bibitem [{\citenamefont {Lorenz}(1872)}]{lorenz1872}%
  \BibitemOpen
  \bibfield  {author} {\bibinfo {author} {\bibfnamefont {L.}~\bibnamefont
  {Lorenz}},\ }\bibfield  {title} {\bibinfo {title} {Determination of the
  degree of heat in absolute measure},\ }\href@noop {} {\bibfield  {journal}
  {\bibinfo  {journal} {Ann. Physik}\ }\textbf {\bibinfo {volume} {147}},\
  \bibinfo {pages} {429} (\bibinfo {year} {1872})}\BibitemShut {NoStop}%
\bibitem [{\citenamefont {Franz}\ and\ \citenamefont
  {Wiedemann}(1853)}]{Franz1853}%
  \BibitemOpen
  \bibfield  {author} {\bibinfo {author} {\bibfnamefont {R.}~\bibnamefont
  {Franz}}\ and\ \bibinfo {author} {\bibfnamefont {G.}~\bibnamefont
  {Wiedemann}},\ }\bibfield  {title} {\bibinfo {title} {Ueber die
  wärme-leitungsfähigkeit der metalle},\ }\href@noop {} {\bibfield  {journal}
  {\bibinfo  {journal} {Annalen der Physik}\ }\textbf {\bibinfo {volume}
  {165}},\ \bibinfo {pages} {497} (\bibinfo {year} {1853})}\BibitemShut
  {NoStop}%
\bibitem [{\citenamefont {Butterworth}\ \emph {et~al.}(2022)\citenamefont
  {Butterworth}, \citenamefont {Triqueneaux}, \citenamefont {Midlik},
  \citenamefont {Golokolenov}, \citenamefont {Gerardin}, \citenamefont
  {Gandit}, \citenamefont {Donnier-Valentin}, \citenamefont {Goupy},
  \citenamefont {Phuthi}, \citenamefont {Schmoranzer}, \citenamefont {Collin},\
  and\ \citenamefont {Fefferman}}]{Butterworth2022}%
  \BibitemOpen
  \bibfield  {author} {\bibinfo {author} {\bibfnamefont {J.}~\bibnamefont
  {Butterworth}}, \bibinfo {author} {\bibfnamefont {S.}~\bibnamefont
  {Triqueneaux}}, \bibinfo {author} {\bibfnamefont {{\v{S}}.}~\bibnamefont
  {Midlik}}, \bibinfo {author} {\bibfnamefont {I.}~\bibnamefont {Golokolenov}},
  \bibinfo {author} {\bibfnamefont {A.}~\bibnamefont {Gerardin}}, \bibinfo
  {author} {\bibfnamefont {T.}~\bibnamefont {Gandit}}, \bibinfo {author}
  {\bibfnamefont {G.}~\bibnamefont {Donnier-Valentin}}, \bibinfo {author}
  {\bibfnamefont {J.}~\bibnamefont {Goupy}}, \bibinfo {author} {\bibfnamefont
  {M.~K.}\ \bibnamefont {Phuthi}}, \bibinfo {author} {\bibfnamefont
  {D.}~\bibnamefont {Schmoranzer}}, \bibinfo {author} {\bibfnamefont
  {E.}~\bibnamefont {Collin}},\ and\ \bibinfo {author} {\bibfnamefont
  {A.}~\bibnamefont {Fefferman}},\ }\bibfield  {title} {\bibinfo {title}
  {Superconducting aluminum heat switch with {3~n$\Omega$} equivalent
  resistance},\ }\href@noop {} {\bibfield  {journal} {\bibinfo  {journal}
  {Review of Scientific Instruments}\ }\textbf {\bibinfo {volume} {93}},\
  \bibinfo {pages} {034901} (\bibinfo {year} {2022})}\BibitemShut {NoStop}%
\bibitem [{\citenamefont {Arenz}\ \emph {et~al.}(1982)\citenamefont {Arenz},
  \citenamefont {Clark},\ and\ \citenamefont {Lawless}}]{Arenz1982}%
  \BibitemOpen
  \bibfield  {author} {\bibinfo {author} {\bibfnamefont {R.~W.}\ \bibnamefont
  {Arenz}}, \bibinfo {author} {\bibfnamefont {C.~F.}\ \bibnamefont {Clark}},\
  and\ \bibinfo {author} {\bibfnamefont {W.~N.}\ \bibnamefont {Lawless}},\
  }\bibfield  {title} {\bibinfo {title} {Thermal conductivity and electrical
  resistivity of copper in intense magnetic fields at low temperatures},\
  }\href@noop {} {\bibfield  {journal} {\bibinfo  {journal} {Phys. Rev. B}\
  }\textbf {\bibinfo {volume} {26}},\ \bibinfo {pages} {2727} (\bibinfo {year}
  {1982})}\BibitemShut {NoStop}%
\bibitem [{\citenamefont {Fickett}(1988)}]{Fickett1988}%
  \BibitemOpen
  \bibfield  {author} {\bibinfo {author} {\bibfnamefont {F.}~\bibnamefont
  {Fickett}},\ }\bibfield  {title} {\bibinfo {title} {Transverse
  magnetoresistance of oxygen-free copper},\ }\href@noop {} {\bibfield
  {journal} {\bibinfo  {journal} {IEEE Transactions on Magnetics}\ }\textbf
  {\bibinfo {volume} {24}},\ \bibinfo {pages} {1156} (\bibinfo {year}
  {1988})}\BibitemShut {NoStop}%
\bibitem [{\citenamefont {Drung}\ \emph {et~al.}(2007)\citenamefont {Drung},
  \citenamefont {Abmann}, \citenamefont {Beyer}, \citenamefont {Kirste},
  \citenamefont {Peters}, \citenamefont {Ruede},\ and\ \citenamefont
  {Schurig}}]{Drung2007}%
  \BibitemOpen
  \bibfield  {author} {\bibinfo {author} {\bibfnamefont {D.}~\bibnamefont
  {Drung}}, \bibinfo {author} {\bibfnamefont {C.}~\bibnamefont {Abmann}},
  \bibinfo {author} {\bibfnamefont {J.}~\bibnamefont {Beyer}}, \bibinfo
  {author} {\bibfnamefont {A.}~\bibnamefont {Kirste}}, \bibinfo {author}
  {\bibfnamefont {M.}~\bibnamefont {Peters}}, \bibinfo {author} {\bibfnamefont
  {F.}~\bibnamefont {Ruede}},\ and\ \bibinfo {author} {\bibfnamefont
  {T.}~\bibnamefont {Schurig}},\ }\bibfield  {title} {\bibinfo {title} {Highly
  sensitive and easy-to-use {SQUID} sensors},\ }\href@noop {} {\bibfield
  {journal} {\bibinfo  {journal} {IEEE Transactions on Applied
  Superconductivity}\ }\textbf {\bibinfo {volume} {17}},\ \bibinfo {pages}
  {699} (\bibinfo {year} {2007})}\BibitemShut {NoStop}%
\bibitem [{\citenamefont {Casey}\ \emph {et~al.}(2021)\citenamefont {Casey},
  \citenamefont {Saunders}, \citenamefont {Levitin},\ and\ \citenamefont
  {van~der Vliet}}]{CSNT2021}%
  \BibitemOpen
  \bibfield  {author} {\bibinfo {author} {\bibfnamefont {A.}~\bibnamefont
  {Casey}}, \bibinfo {author} {\bibfnamefont {J.}~\bibnamefont {Saunders}},
  \bibinfo {author} {\bibfnamefont {L.}~\bibnamefont {Levitin}},\ and\ \bibinfo
  {author} {\bibfnamefont {H.}~\bibnamefont {van~der Vliet}},\ }\href@noop {}
  {\bibinfo {title} {Current sensing noise thermometer}} (\bibinfo {year}
  {2021}),\ \bibinfo {note} {{P}atent International Publication number: WO
  2021/239513 A1}\BibitemShut {NoStop}%
\end{thebibliography}%
\end{document}